\documentclass[11pt,a4paper]{article}
\pdfoutput=1
\usepackage{jheppubmod}
\usepackage[utf8]{inputenc}
\usepackage{epstopdf}
\usepackage{graphicx}
\usepackage{epsfig}
\usepackage{dcolumn}  
\usepackage{bm}    
\usepackage{caption}
\usepackage{subcaption}
\usepackage{amssymb} 
\usepackage{amsmath,bm}
\usepackage{amsfonts}  
\usepackage{amsmath}  
\usepackage{slashed}  
\usepackage{enumitem}
\usepackage[mathscr]{euscript}
\usepackage{tabu}
\usepackage{epsfig}
\makeatletter
\g@addto@macro\bfseries{\boldmath}
\makeatother

\def\l1{{{1-loop}}}

\def\n1{\Bigg|_{n=1}}

\def\n{{(n)}}

\usepackage[T1]{fontenc} 
\usepackage{tikz}
\usepackage{amsmath,amssymb}
\usepackage{relsize}
\usepackage{latexsym}
\usepackage{leftidx}
\usepackage{diagbox}
\usepackage[T1]{fontenc}
\usepackage{array}
\usepackage{makecell}
\usepackage{csquotes}
\usepackage{tikz}
\usepackage{enumitem}
\usepackage{setspace}
\usepackage{multirow}
\usepackage{amsmath,amssymb}
\usepackage{relsize}
\usepackage{latexsym}
\usepackage{leftidx}
\usepackage{csquotes}
\usepackage{tikz}
\usepackage{enumitem}
\usetikzlibrary{decorations.markings}
\usetikzlibrary{decorations.pathmorphing}
\usetikzlibrary{decorations.markings}
\usetikzlibrary{decorations.pathmorphing}
\usepackage{pifont}

\usepackage{bookmark}
  \title{\textbf{\textsf{Partition functions of higher derivative conformal fields on conformally related spaces}}}

\author{ Jyotirmoy Mukherjee}
\affiliation{\vspace{.1cm} Centre for High Energy Physics, \\ Indian Institute of Science,\\
C. V. Raman Avenue, Bangalore 560012, India.}
\emailAdd{ jyotirmoym@iisc.ac.in}
\abstract{The character integral representation of one loop partition functions is useful to establish the relation between partition functions of conformal fields on Weyl equivalent spaces. The Euclidean space $S^a\times AdS_b$ can be mapped to $S^{a+b}$ provided $S^a$ and $AdS_b$ are of the same radius. As an example, to begin with, we show that the partition function in the character integral representation of conformally coupled free scalars and fermions are identical on  $S^a\times AdS_b$  and  $S^{a+b}$. We then demonstrate that the partition function of higher derivative conformal scalars and fermions are also the same on hyperbolic cylinders and branched spheres.  The partition function of the four-derivative conformal vector gauge field on the branched sphere in $d=6$ dimension can be expressed as an integral over `naive' bulk and  `naive' edge characters. However, the partition function of the conformal vector gauge field on $S^1_q\times AdS_5$ contains only the `naive' bulk part of the partition function. This follows the same pattern which was observed for the partition of conformal $p$-form fields on hyperbolic cylinders. We use the partition function of higher derivative conformal fields on hyperbolic cylinders to obtain a linear relationship between the Hofman-Maldacena variables which enables us to show that these theories are non-unitary.  }
\begin{document}

\maketitle

\section{Introduction}
The one-loop partition function of free fields is an important ingredient to evaluate entanglement entropies, anomaly coefficients in even dimensions, $F$-terms in odd dimensions. Partition functions on hyperbolic cylinders and branched spheres are useful to evaluate R\'{e}nyi entropies and entanglement entropies of conformal fields across a sphere  \cite{Herzog:2015ioa, Hung:2014npa}. Therefore it receives attention in the area of condensed matter physics \cite{Calabrese:2007mtj, Calabrese:2009qy} as well as in quantum gravity. \\
It is quite well known that the hyperbolic cylinder is related to the branched sphere by Weyl transformation. Moreover the Euclidean space $S^a\times AdS_b$ can be mapped to $S^{a+b}$ where $S^a$ is $a$-dimensional sphere and  $AdS_b$ is $b$-dimensional anti-de Sitter space. The conformal map works only when the sphere and the anti-de Sitter space have the same radius. To understand the  Weyl transformation we write the metric of $S^{a+b}$
\begin{align}
    ds^2_{S^{a+b}}=d\tau^2+\cos^2\tau ds^2_{S^a}+\sin^2\tau ds^2_{S^{b-1}}.
\end{align}
Now one uses the following transformation
$$\tan\tau=\sinh u,$$ and obtains
\begin{align}\label{metric}
     ds^2_{S^{a+b}}&=\frac{1}{\cosh^2 u}\left(ds^2_{S^a}+du^2+\sinh^2uds^2_{S^{b-1}}\right).
\end{align}
One can identify \eqref{metric} with the metric of $S^a\times AdS_b$ barring the conformal factor $\frac{1}{\cosh^2 u}$.
Since these euclidean spaces are related by Weyl transformation one expects that partition functions of the conformal fields will be identical on these backgrounds. Similar kinds of questions have been addressed before \cite{Rodriguez-Gomez:2017kxf} where  $F$-terms of the conformal free scalar fields were compared in odd dimensions. In \cite{Klebanov:2011uf} it was explicitly verified that the partition functions of conformally coupled scalars and fermions on $ S^1\times AdS_2$ and on $S^3$ coincide.

Recently in \cite{Anninos:2020hfj, Sun:2020ame} it was shown that partition functions of scalars, fermions, and integer higher spin fields can be expressed as integral over characters on a euclidean patch of de Sitter as well as on anti-de Sitter spaces. In \cite{David:2021wrw}  partition functions of the anti-symmetric $p$-form fields on sphere and anti-de Sitter space were expressed in terms of the bulk and the edge characters where the edge character of a co-exact $p$-form contains characters of anti-symmetric tensors of rank lower to $p$ all the way to the zero-form. It was also shown that for conformal $p$-forms the hyperbolic cylinder does not capture the edge mode part of the partition function.
We summarise the work of \cite{Anninos:2020hfj,Sun:2020ame,David:2021wrw} briefly where the one loop partition function of  scalars, fermions, integer higher spin fields and anti-symmetric $p$-form fields were expressed in terms of the integral over the `bulk' and the `edge' part of the Harish-Chandra character.
\begin{align}
    \log\mathcal{Z}&=\int_0^{\infty}\frac{dt}{2t}
    \begin{cases}
        \frac{ 1+ e^{-t} }{ 1-  e^{-t} }
 \left( \chi_{{\rm bulk }} ( t) - \chi_{{\rm edge}} ( t) 
 \right) \qquad \rm{bosonic \hspace{1mm} field}, \\
  \frac{ 2e^{-\frac{t}{2}} }{ 1-  e^{-t} }       
 \left( \chi_{{\rm bulk }} ( t) - \chi_{{\rm edge}} ( t) 
 \right) \qquad \rm{fermionic\hspace{1mm} field}.
    \end{cases}
\end{align}
Note that the kinematic factor is different in bosonic and fermionic fields.\\
Here we only consider the character part of the partition function but there is another part that depends on the dimensionless coupling constant of the theory and the volumes of the gauge group of the fields \cite{Anninos:2020hfj}. We will be interested only in the character part of the partition function because it contains anomaly coefficients in even dimensions and the $F$-terms in odd dimensions.

We find that the partition function of the conformally coupled free scalar and free fermion on $S^a\times AdS_b$ can be expressed in terms of integral over characters. We demonstrate that for conformally coupled free scalars as well as for free fermions one can sum over the eigenmodes on $S^a$  and show that the character which to begin with was an $AdS$ character indeed becomes the character on the $S^{a+b}$.  Therefore it establishes the fact that partition functions of free conformal scalar and fermions are identical in the Weyl equivalent spaces in arbitrary dimensions. This was checked for scalars explicitly \cite{Rodriguez-Gomez:2017kxf} in few cases of $a$ and $b$ and recently by \cite{Nishioka:2021uef} for $a+b=100$ numerically. For $a=1$, $S^a\times AdS_b$ corresponds to the hyperbolic cylinder and one can evaluate the entanglement entropy across a sphere by mapping it to the hyperbolic cylinder and evaluate the thermodynamic entropy on it. One can also evaluate the conformal dimension of the twist operator which essentially captures the energy density in the presence of the defect across the entangling surface.
\begin{align}
    \begin{split}
         S_q&=\frac{-\mathcal{F}_q+q\mathcal{F}_{q=1}}{1-q}, \qquad\qquad
          S_{\rm{EE}}=\lim_{q\rightarrow 1}S_q,\\
 h_q&=\frac{q}{(d-1)\rm{Vol}(AdS_{d-1})}\left(\partial_q\mathcal{F}_q|_{q=1}-\partial_q\mathcal{F}_q\right).
    \end{split}
\end{align}
Here $\mathcal{F}_q$ is the free energy on hyperbolic cylinder. The regularised volume of $AdS_{d}$  is  given by \cite{Hung:2014npa}
\begin{eqnarray}\label{volads}
{\rm Vol } ( AdS_{d}) = 
\begin{cases}
\pi^{\frac{d-1}{2} } \Gamma(  - \frac{d-1}{2} )  \qquad  & \hbox{for}\;  d  \; \hbox{even} ,  \\
\frac{ 2 ( -\pi)^{\frac{d-1}{2}  }}{ \Gamma( \frac{d+1}{2} ) } \log  R
\qquad  & \hbox{for} \; d\; \hbox{odd}.
\end{cases}
\end{eqnarray}
Here $R$ is the ratio of the radial cutoff on $AdS$ to the radius of $AdS$.

The higher derivative  Weyl-invariant theory has been studied extensively in \cite{Beccaria:2014jxa, Beccaria:2017dmw, Beccaria:2017lcz} to understand the conformal anomaly of classically Weyl invariant theories. In this paper, we study partition functions of the higher derivative Weyl invariant theories on hyperbolic cylinders as well as on branched spheres. 
  We use the method of factorization \cite{Beccaria:2017dmw} to write the higher derivative operator in terms of the product of two-derivative operators. Therefore the free energy becomes the sum over all the free energies of factorized two derivative operators.
To understand this let  us consider a Weyl invariant $2r$-derivative operator  which factorizes into $r$ number of $2$-derivative operators in the following way
\begin{align}
    \mathcal{O}^{~2r}&=\prod_{i=1}^{r}\left(-\nabla^2+m^2_{(i)}\right),
\end{align}
where $\nabla^2$ is the two-derivative  operator and $m_{(i)}$ corresponds to the mass of each two-derivative operator which comes from the curvature coupling. We demonstrate that the partition function of higher derivative fields can be written as integral over characters where the character of the $2r$-derivative operator is the sum of $r$ number of $2$-derivative operators.
\begin{align}
  -\mathcal{F}^{2r}=\int_0^{\infty}\frac{dt}{2t}
  \begin{cases}
       \frac{ 1+ e^{-t} }{ 1-  e^{-t} }
\sum_{i=1}^{r} \left( \chi_{{\rm bulk },i} ( t) - \chi_{{\rm edge},i} ( t) 
 \right) \qquad \rm{bosonic \hspace{1mm} field}, \\
  \frac{ 2e^{-\frac{t}{2}} }{ 1-  e^{-t} }       
\sum_{i=1}^{r} \left( \chi_{{\rm bulk },i} ( t) - \chi_{{\rm edge},i} ( t) 
 \right) \qquad \rm{fermionic\hspace{1mm} field}.
  \end{cases}
\end{align}
Note that this factorization works even if the two derivative kinetic operators admit zero-mode. Generally, one excludes the zero-mode which essentially gives a shift in the expression of free energy but for our purpose, we will keep it because it helps to cast the partition function in terms of characters. 

We study Weyl invariant $4$-derivative scalar field on the hyperbolic cylinder and express the partition function in terms of the integral over characters. We first factorize the $4$-derivative operator in terms of two 2-derivative operators. We perform sum over  Kaluza -Klein modes on $S^1_q$  and show that the character which to begin with was an $AdS$ character indeed becomes the character on the $S^{d}_q$. Now one can evaluate entanglement entropy and the weight of the twist operator of $4$-derivative conformal scalar field across a spherical entangling surface from the free energy on the hyperbolic cylinder or branched sphere. We also study the Weyl invariant $6$-derivative scalar field and show that the partition function can be expressed as integral over characters and the integrands are identical on the hyperbolic cylinder and on the branched sphere. 

It is well known that massless spin-$1$ or free Maxwell field is conformal in $d=4$ dimension.  Therefore one expects the partition function will be identical on $S^4_q$ and $AdS_3\times S^1_q$.
In \cite{Anninos:2020hfj} it was shown that the partition function of complete spin-1 or $1$-form field can be expressed in terms of the bulk and edge Harish Chandra character. In
\cite{David:2021wrw} partition function of conformal $p$-forms were expressed in terms of the integrals over character on the branched sphere as well as on hyperbolic cylinder. It was explicitly shown that the hyperbolic cylinder does not capture the edge mode part of the partition function of the $1$-form field. In this paper, we express the partition function of $4$-derivative Weyl invariant vector gauge field in terms of the `naive' bulk and `naive' edge characters on $S^6_q$. 
\begin{eqnarray}\label{vectorsphereintro}
\log\mathcal{Z}^{(4)}_v&=&\int_0^{\infty}\frac{dt}{2t}\left\{ 
    \frac{1+e^{-\frac{t}{q} } }{1-e^{-\frac{t}{q}}}\hat{\chi}_{\rm{bulk}}- \frac{1+e^{-t} }{1-e^{-t} }\hat{\chi}_{\rm{edge}}\right\},
\end{eqnarray}
where $\hat{\chi}_{\rm{bulk}}$ and $\hat{\chi}_{\rm{edge}}$ are the `naive' bulk and `naive' edge characters in the sense of \cite{Anninos:2020hfj} and they do not belong to the UIR of any exceptional series. We  evaluate the partition function of the $4$-derivative Weyl invariant vector gauge field on $AdS_5\times S^1_q$ by summing over Kaluza-Klein modes along $S^1_q$ direction and obtain a character integral which agrees with only  the `naive' bulk character of \eqref{vectorsphereintro} and misses out the `naive' edge part. Therefore it demonstrates that hyperbolic cylinder does not capture the edge part of the character integral in partition function. This supports the earlier observations \cite{Huang:2014pfa,Donnelly:2014fua,Donnelly:2015hxa,Nian:2015xky,Dowker:2017flz,David:2020mls} found in the context of evaluation of  entanglement entropy of conformal $p$-form fields.

Theories with the higher derivative kinetic terms in the action admit negative residue in the propagator. This implies negative norm states and therefore these theories are expected to be non-unitary\cite{Donoghue:2017fvm}. 
To understand the non-unitary nature of higher derivative conformal fields we use the causality bound on the Hofman-Maldacena variables $t_2$ and $t_4$ \cite{Camanho:2009vw} which comes from the positivity of the energy flux. The variables $t_2$, $t_4$ were constructed from the ratio of the linear combination of the three parameters $a$, $b$ and $c$ determining the two and three-point functions of the stress tensors \cite{Osborn:1993cr}. The bounds on the parameters represent a triangle in $t_2-t_4$-plane. In \cite{Hung:2014npa} it was shown that the first and second derivative of the conformal dimension of the twist operator can be expressed in terms of two and three-point functions of the stress tensor. Therefore we obtain two relations involving the parameters $a$, $b$, and $c$  which do not fully determine the values of the parameters. However, we find a linear relation of the variables $t_2$, $t_4$ which represents a straight line in the plane, and the theory is now constrained to lie on this straight line. We observe that the straight line never intersects the region of causality bound which implies that the higher derivative Weyl invariant theory which is just a point on the straight line can never lie inside the region of causality bound. Therefore we conclude that the higher derivative Weyl invariant theories do not satisfy the causality bound. 
\section{Conformal Scalar}
In this section we study the one-loop partition function of Weyl-invariant scalar on $S^d$ and $S^a\times AdS_b$ where $S^a$ is $a$-dimensional sphere and $AdS_b$ is the $b$-dimensional anti-de Sitter space. We show that that the partition functions are identical when $a+b=d$.  We express the partition functions in terms of integrals over the Harish-Chandra character. The partition function on the space $S^a\times AdS_b$ is useful when $a=1$ and one can evaluate entanglement entropy and the conformal dimension of the twist operator of conformal field theories across a spherical entangling surface \cite{Casini:2011kv}. We verify that the log divergence of the partitions functions agrees with previous evaluations by \cite{Nishioka:2021uef}. The work of \cite{Anninos:2020hfj} and \cite{Sun:2020ame} have shown that the one loop partition functions on the sphere and anti-de Sitter space have a nice character representation. In the latest work of \cite{David:2021wrw} has demonstrated the equality of the partition function of conformal fields on $S^{d+1}$ and $S^1\times AdS_d$. Therefore it is a natural question to ask the same can be generalized to the arbitrary family of spaces of the form $S^a\times AdS_b$.\\
\begin{subsection}{Conformal scalar on \texorpdfstring{$S^d$}{TEXT} }To explain the techniques in detail we begin with evaluating the partition function of conformally coupled free scalar field on $S^d$.
The partition function of the conformal field on $S^d$ can be written in terms of integrals over the Harish-Chandra character \cite{Anninos:2020hfj, David:2021wrw}.
The Weyl invariant action of the real scalar  in $d$ dimensions is given by 
\begin{equation}\label{action}
S =- \frac{1}{2}  \int d^{d} x \sqrt{g} (  \partial_\mu\phi \partial^\mu \phi   + \frac{d-2}{4 (d-1)} R \phi^2) .
\end{equation}
The curvature on of $S^d$ is known to be $R=d(d-1)$ . Therefore the partition function of a conformal scalar on $S^d$ can be written as
 \begin{eqnarray}
     { \mathcal Z}[S^d ]^{(0)} =\frac{1}{ {\rm det} ( -\Delta^{S^d}_0+m_{S^d}^2 )^{\frac{1}{2} } } .
 \end{eqnarray}
 Where $\Delta^{S^d}_0$ is the spin-0 Laplacian on $S^d$.
 The curvature induced mass turns out to be $m^2_{S^d}=\frac{d(d-2)}{4}$.
 The eigen-value and corresponding degeneracy of scalar Laplacian on $S^d$ are known \cite{Camporesi:1994ga}
 \begin{align}\label{spheredeg}
     \begin{split}
      \lambda_n^{0}&=n(n+d-1)\\
      g^{0}_n&=\frac{(d+2 n-1) \Gamma (d+n-1)}{n! \Gamma (d)}\qquad  n  \in \{ 0, \cdots \infty \},
     \end{split}
 \end{align}
 Therefore the free energy of the conformal scalar on $S^d$ becomes
 \begin{eqnarray}\label{scalar partition}
    \log\mathcal{Z}[S^{d}]^{(0)} &=& - \frac{1}{2} 
    \sum_{n=0}^\infty  g_{n}^{(0)} \log \left( \lambda_{n}^{(0)} +m^2_{S^d}  \right) ,
    \\ \nonumber
    &=& \int_0^\infty \frac{d\tau}{2\tau}  \sum_{n =0}^\infty
    g_{n }^{(0)} (  e^{ - \tau ( \lambda_{n}^{(0)}+m^2_{S^d} )}  - e^{-\tau} ) .
    \end{eqnarray}
    In the second line we have used the identity \begin{equation}\label{logiden}
-\log y  = \int_0^\infty \frac{d\tau}{\tau} ( e^{-y \tau} - e^{-\tau} ) .
\end{equation}
We need to compute the sum over degeneracy in order to evaluate the second term in the partition function. We observe that in the large $n$ limit, the degeneracy $g_n^{(0)}$ converges in the sufficiently large negative value of d.
    \begin{align}
        g_n^{(0)}\sim \left(\frac{1}{n}\right)^{-d+1} \left(\frac{2 (d-3) (d-4)!}{(d-3)! (d-1)! }+\mathcal{O}\left(\frac{1}{n}\right)\right).
    \end{align}
    Therefore we can perform the sum by using zeta regularisation prescription and obtain\cite{Klebanov:2011gs}
    \begin{equation} \label{Hubbard}
    \sum_{n=0}^\infty g_{n }^{(0)}  =0.
    \end{equation} 
    So the second term in \eqref{scalar partition} vanishes and we proceed with the first term 
     \begin{equation}\label{4Dpart}
        \log\mathcal{Z}[S^{d}]^{(0)} =   \int_{\epsilon}^\infty \frac{d\tau}{2\tau} e^{-\frac{\epsilon^2}{4\tau} } 
       \sum_{n =0}^\infty g_{n }^{(0)} e^{ - \tau\left( ( n+\frac{d-1}{2} )^2-\frac{1}{4}\right)}  .
    \end{equation}
It is convenient to introduce the factor  $e^{-\frac{\epsilon^2}{4\tau} } $ at this point which will help us to keep track of the branch cut in the integration plane. Note that this does not serve as a regulator in this case as the starting integral is already convergent \eqref{scalar partition} after dropping the second term by using the dimension regularisation \cite{David:2021wrw}.  \\  We now use Hubbard-Stratonovich trick to perform the sum over the eigen modes on sphere
    \begin{equation} \label{contint}
   \sum_{n=0}^{\infty}g_{n}^{(0)} e^{-\tau(n+\frac{d-1}{2})^2}
   =\int_{C} \frac{du}{\sqrt{4\pi\tau}}e^{-\frac{u^2}{4\tau}}f_0(u).
\end{equation}
Here the contour $C$ runs from $-\infty $ to $\infty$ slightly above the real axis and $f_0(u)$ is given by
\begin{align}
   f_0(u) &=\sum_{n=0}^{\infty}g_{n}^{(0)} e^{i u(n+\frac{d-1}{2})}\nonumber\\
&=\frac{e^{\frac{(d-1) i u}{2}} \left(1+e^{i u}\right)}{\left(-1+e^{\frac{i u}{2}}\right)^d \left(1+e^{\frac{i u}{2}}\right)^d}. 
\end{align}
We  perform the  integral over $\tau$ in \eqref{4Dpart} which results in 
\begin{equation}\label{u part}
 \log\mathcal{Z}[S^{d}]^{(0)}  =
\int_{C}\frac{du}{ 2\sqrt{u^2+\epsilon^2}}\left(e^{-\nu_0\sqrt{u^2+\epsilon^2}}f_0(u)\right).
\end{equation}
Now one can deform the contour 
$C$ from the real line to the contour  $C'$ which runs on the both sides of the branch cut on the imaginary axis 
  which originates at $u = i \epsilon$ on the  $u$-plane. This is shown in figure [\ref{fig1}].
Substituting $u = i t$ we obtain
\begin{equation} 
\log\mathcal{Z}[S^{d}]^{(0)}  =
\int_{\epsilon}^{\infty}\frac{dt}{2\sqrt{t^2-\epsilon^2}}\left(e^{i\nu_0\sqrt{t^2-\epsilon^2}}+e^{-i\nu_0\sqrt{t^2-\epsilon^2}}\right)f_0(i t).
\end{equation}
We can now take $\epsilon\rightarrow 0$ and the free energy becomes
\begin{equation} \label{endstep}
\log\mathcal{Z}[S^{d}]^{(0)}  =
\int_{0}^{\infty}\frac{dt}{2t}\frac{1+e^{-t}}{1-e^{-t}}
  \chi_{ (d, 0)\;{\rm conf}  }^{dS} ( t) ,
\end{equation}
 where  the  $\rm{SO}(1, d)$ Harish-Chandra character   is given by \cite{Anninos:2020hfj}
  \begin{equation} \label{charconfsc}
   \chi_{(d, 0)\;{\rm conf}  }^{dS} ( t)  = 
   \frac{ e^{ - \frac{(d-2)}{2} t } + e^{ - \frac{d}{2} t } }{ ( 1- e^{-t} )^{d-1}} .
  \end{equation}

\begin{figure}[h]
\centering
\begin{tikzpicture}[thick,scale=0.85]
\filldraw[magenta] 
                (0,0.5) circle[radius=3pt]
                (0,-0.5) circle[radius=3pt];
                \filldraw[orange] 
                (0.1,0) circle[radius=2pt]
                (0.5,0) circle[radius=2pt]
              (1,0) circle[radius=2pt]  
              (1.5,0) circle[radius=2pt]
              (2,0) circle[radius=2pt]
               (2.5,0) circle[radius=2pt]
                (3,0) circle[radius=2pt]
               (3.5,0) circle[radius=2pt]
                (4,0) circle[radius=2pt]
                (-0.1,0) circle[radius=2pt]
                (-0.5,0) circle[radius=2pt]
              (-1,0) circle[radius=2pt]  
              (-1.5,0) circle[radius=2pt]
              (-2,0) circle[radius=2pt]
               (-2.5,0) circle[radius=2pt]
                (-3,0) circle[radius=2pt]
               (-3.5,0) circle[radius=2pt]
                (-4,0) circle[radius=2pt] ;
\draw [decorate,decoration=snake] (0,-0.5) -- (0,-4);
\draw [decorate,decoration=snake] (0,0.5) -- (0,4);
\draw [postaction={decorate,decoration={markings , 
mark=at position 0.55 with {\arrow[black,line width=0.5mm]{<};}}}](0.2,1) arc[start angle=0, end angle=-180, radius=0.2cm];
\draw
[
postaction={decorate,decoration={markings , 
mark=at position 0.20 with {\arrow[red,line width=0.5mm]{>};}}}
][red, thick] (-4,0.2)--(0,0.2);
\draw[red, thick] (0,0.2)--(4,0.2);
\draw (0.4,1) node{$\mathbf{\epsilon}$};
\draw (0.4,-1) node{$\mathbf{-\epsilon}$};
\draw[gray, thick] (0,0) -- (0,4);
\draw[gray, thick] (0,0) -- (0,-4);
\draw[gray, thick] (0.2,4) -- (0.2,1);
\draw[gray, thick] (-0.2,1) -- (-0.2,4);
\draw[gray, thick] (0,0) -- (4,0);
\draw[gray, thick] (0,0) -- (-4,0);
\draw[gray, thick] (2,3) -- (2,3.4);
\draw[gray, thick] (2,3) -- (2.5,3);
\draw (2.3,3.3) node{$\mathbf{u}$};
\draw [red,thick](-2,0.5) node{$\mathbf{C}$};
\draw [gray,thick](0.5,3) node{$\mathbf{C'}$};
\end{tikzpicture}
\caption{Integration contour in $u$-plane} \label{fig1}
\qquad
\centering
\begin{tikzpicture}[thick,scale=0.85]
\filldraw[magenta] 
                (0.5,0) circle[radius=3pt]
                (-0.5,0) circle[radius=3pt];
                \filldraw[orange] 
               ( 0,0.2) circle[radius=2pt]
                (0,0.5) circle[radius=2pt]
              (0,1) circle[radius=2pt]  
              (0,1.5) circle[radius=2pt] 
              (0,2) circle[radius=2pt] 
              (0,2.5) circle[radius=2pt]
                (0,3) circle[radius=2pt]
               (0,3.5) circle[radius=2pt]
                (0,4) circle[radius=2pt]
                (0,-0.1) circle[radius=2pt]
                (0,-0.5) circle[radius=2pt]
              (0,-1) circle[radius=2pt]  
              (0,-1.5) circle[radius=2pt]
              (0,-2) circle[radius=2pt]
               (0,-2.5) circle[radius=2pt]
                (0,-3) circle[radius=2pt]
               (0,-3.5) circle[radius=2pt]
                (0,-4) circle[radius=2pt] ;
\draw [decorate,decoration=snake] (-0.5,0) -- (-4,0);
\draw [decorate,decoration=snake] (0.5,0) -- (4,0);
\draw
[
postaction={decorate,decoration={markings , 
mark=at position 0.20 with {\arrow[red,line width=0.5mm]{>};}}}
][blue, thick] (0.5,0.2)--(4,0.2);
\draw (1,0.4) node{$\mathbf{\epsilon}$};
\draw (-1,0.4) node{$\mathbf{-\epsilon}$};
\draw[gray, thick] (0,0) -- (0,4);
\draw[gray, thick] (0,0) -- (0,-4);
\draw[gray, thick] (0,0) -- (4,0);
\draw[gray, thick] (-4,0) -- (0,0);
\draw[gray, thick] (2,3) -- (2,3.4);
\draw[gray, thick] (2,3) -- (2.5,3);
\draw (2.3,3.3) node{$\mathbf{t}$};
\end{tikzpicture}
\caption{Integration contour in $t$-plane} \label{fig2}
\end{figure}

The expression of $\chi_{(d, 0)\;{\rm conf}  }^{dS} ( t)$ agrees with the character of the scalar field on $S^d$ \cite{Anninos:2020hfj} with the substitution of $\nu_0=\frac{i}{2}$ and the partition function coincides with the partition function of conformal scalar field on branched sphere $S^d_q$ with $q=1$ given in \cite{David:2021wrw}.
\end{subsection}
\begin{subsection}{Conformal Scalar on \texorpdfstring{$S^a\times AdS_b$}{TEXT}}
 The action of a conformally coupled scalar on $S^a\times AdS_b$ is given in \eqref{action}
where $d=a+b.$ The free energy can be written as
\begin{align}
  - \log \mathcal{Z}[S^a\times AdS_{b}]^{(0)}&=\frac{1}{2}\rm{Tr}\log(-\Delta_0+m_0^2),
\end{align}
where $\Delta_0$ is the scalar Laplacian on $S^a\times AdS_b$ which can be decomposed into the spin-0 Laplacian on $S^a$ and the spin-0 Laplacian on $AdS_b$. $$-\Delta_0=-\Delta_a-\Delta_b.$$ $\Delta_a$ is the scalar Laplacian on $S^a$ with the eigen-value and degeneracy is given in \eqref{spheredeg} and
$\Delta_b$ is scalar Laplacian on $AdS_b$. The eigen values of  the spin-0 Laplacian  on $AdS_b$  are given by \cite{Camporesi:1994ga} 
\begin{equation}\label{eigen}
 \Delta_{b} \psi^{\{\lambda, u\}} = -\left[  \lambda^2 + (  \frac{b-1}{2} )^2 \right] 
 \psi_\lambda^{\{\lambda, u\}} ,
 \end{equation}
 $\psi^{ \{\lambda, u\} }$  are the corresponding eigen functions and $\{u\}$ labels other quantum numbers
 on $AdS_b$.
The curvature induced mass $m_0^2$ on $S^a\times AdS_b$ becomes
\begin{align}
    m_0^2=\frac{(a-b)(a+b-2)}{4}.
\end{align}
Using these eigen values  and the degeneracies
we evaluate the free energy
\begin{align}
     - \log \mathcal{Z}[S^a\times AdS_{b}]^{(0)}&=\frac{1}{4}\sum_{n=0}^{\infty}g^{(0)}_n\int_{-\infty}^{\infty}\mu^{(0)}_b(\lambda)\log\left(\lambda^2+(n+\frac{a-1}{2})^2\right).
\end{align}
Note that shift in the eigen value of $AdS_b$ together with factor $\frac{(a-1)^2}{4}$ from the eigen value of the spin zero Laplacian on $S^a$ precisely cancel the mass term. Now we use the identity \eqref{logiden} to replace the logarithm
\begin{align}
     \log \mathcal{Z}[S^a\times AdS_{b}]^{(0)}&=\frac{1}{4} \int_0^\infty \frac{d\tau}{\tau} \sum_{n=0}^\infty g^{(0)}_n
  \int_{-\infty}^{\infty} d\lambda 
  \mu_{0}^{(d)}(\lambda)  ( e^{ -\tau( \lambda^2 +(n+\frac{a-1}{2})^2 ) } - e^{-\tau} ).
\end{align}
Using the same logic presented in \eqref{Hubbard} we drop the second term where we perform the sum over degeneracy using zeta regularisation prescription and therefore proceed with the first term to obtain
\begin{align}
  \log \mathcal{Z}[S^a\times AdS_{b}]^{(0)}&=\int_0^{\infty}\frac{d\tau}{4\tau}e^{-\frac{\epsilon^2}{4\tau}}\sum_{n=0}^{\infty}g^{(0)}_n 
 \int_{-\infty}^{\infty} d\lambda 
  \mu^{(b)}_0 (\lambda)  \left(  e^{ - \tau (\lambda^2+(n+\frac{a-1}{2})^2}  \right)  .  
\end{align}
We perform the integral over $\lambda$ by using Hubbard-Stratonovich  trick
\begin{align}
     \log \mathcal{Z}[S^a\times AdS_{b}]^{(0)}&=\frac{1}{4}\int_{-\infty}^{\infty}d\tau\int_{C}\frac{du}{\sqrt{4\pi\tau^3}} \sum_{n=0}^{\infty}g^{(0)}_ne^{-\frac{\epsilon^2+u^2}{4\tau}}e^{-\tau(n+\frac{a-1}{2})^2}W_0^{(b)}(u),
\end{align}
$W_0^{(b)}$ is the Fourier transform of the Plancherel  measure of spin-0 field on $AdS_b$ \cite{Sun:2020ame}.
\begin{equation}\label{PlFourier}
W_0^{(b)}( u) = \frac{1+ e^{-u} }{ 1- e^{-u}} \frac{ e^{-\frac{b-1}{2} u} }{ ( 1- e^{-u} )^{b-1}} .
\end{equation}
Here the contour  $\rm{C_{odd}}$ or $\rm{C_{even}}$ is given in figure [\ref{fig4}] and figure [\ref{fig5}]
 depending on whether $b$ is odd or even respectively. 
 
 For the case when $b$ is even substituting $W_0^{b}$  and using the 
 contour $\rm{C_{even}}$ as shown in figure [\ref{fig4}] .
We perform the integral over $\tau$ and obtain
\begin{align}\label{KKsum}
        \log \mathcal{Z}[S^a\times AdS_{b}]^{(0)}&=\int_{\rm{C_{even}}}\frac{du}{4\sqrt{u^2+\epsilon ^2}}\sum_{n=0}^{\infty}g^{(0)}_ne^{  \left(-(n+\frac{a-1}{2})\sqrt{u^2+\epsilon ^2}\right)}W_0^{(b)}(u).
\end{align}
 We now  perform the sum over the eigen modes on $S^a$ and take $\epsilon\rightarrow 0$ limit to obtain
\begin{align}\label{uintegralads}
    \log \mathcal{Z}[S^a\times AdS_{b}]^{(0)}&=\frac{1}{2}\int_{R+i\delta}\frac{du}{2u}\frac{1+ e^{-u} }{ 1- e^{-u}}\frac{e^{-u(\frac{a-1}{2})}+e^{-u(\frac{a+1}{2})}}{(1-e^{-u})^a}\frac{ e^{-\frac{b-1}{2} u} }{ ( 1- e^{-u} )^{b-1}} \nonumber\\
      &=\frac{1}{2}\int_{R+i\delta}\frac{du}{2u}\frac{1+ e^{-u} }{ 1- e^{-u}}\frac{e^{-u(\frac{d}{2}-1)}+e^{-u \frac{d}{2}}}{(1-e^{-u})^{d-1}}.
\end{align}
 In the last line we have substituted $a+b=d$. Now we take $\delta\rightarrow 0$ write the partition function as
\begin{align}
   \log \mathcal{Z}[S^a\times AdS_{b}]^{(0)}&= \frac{1}{2}\int_0^{\infty} \frac{du}{2u}\frac{1+ e^{-u} }{ 1- e^{-u}} \chi_{ (d, 0)\;{\rm conf}  }^{dS} ( u) ,
\end{align}
We proceed similarly in the odd $AdS_b$. The contour is given in [\ref{fig5}] and obtain
\begin{align}
    \log \mathcal{Z}[S^a\times AdS_{b}]^{(0)}&=\frac{\log R}{2\pi i}\int_{\rm{C_{odd}}}\frac{du}{2u}\frac{1+ e^{-u} }{ 1- e^{-u}}\frac{e^{-u(\frac{d}{2}-1)}+e^{-u\frac{ d}{2}}}{(1-e^{-u})^{d-1}}.
\end{align}
Note that the partition function of conformal scalar on $S^a\times AdS_b$ is identical to the partition function on $S^{a+b}$.
The character integral representations make the relations  between partition function on conformally related spaces
explicit. 
The partition of branched sphere has $q$ dependent  kinematic factor but the character remains same \cite{David:2021wrw}. One can replace the kinematic factor $\frac{1+ e^{-u} }{ 1- e^{-u}}$ by $\frac{1+ e^{-\frac{u}{q}} }{ 1- e^{-\frac{u}{q}}}$ to obtain the partition function on the branched sphere. \\
We evaluate the log divergent part of the free energy given in \eqref{endstep} of conformal scalar  in even dimension by expanding the integrand around $t=0$ and collecting the coefficient of $\frac{1}{t}$ term. Here we present the log divergent term in even dimension $4\leq d\leq 14$ which agree with \cite{Nishioka:2021uef,David:2021wrw}.
\begin{table}[ht]
\begin{center}
\begin{tabular}{  l | l | l | l | l | l | l |  }
\hline
$d$  &   4    &    6   &   8   &   10   &   12   &   14   \\
\hline
$\log$ divergent term & $-\frac{1}{90}$ & $\frac{1}{756}$ & $-\frac{23}{113400}$ & $\frac{263}{7484400}$ & $-\frac{133787}{20432412000}$ & $\frac{157009}{122594472000}$ \\
\hline
\end{tabular}
\caption{Log divergent term of the patition function of conformal scalar in even dimnsion.}
\end{center}
\end{table}
\begin{figure}[h]
\centering
\begin{tikzpicture}[thick,scale=0.85]
\filldraw[magenta] 
                (0,1) circle[radius=3pt]
                (0,-1) circle[radius=3pt];
\draw [decorate,decoration=snake] (0,-1) -- (0,-4);
\draw [decorate,decoration=snake] (0,1) -- (0,4);
\draw[gray,thick] (-2,0.8) node{$\mathbf{\rm{C_{even}}}$};
\draw (0.4,1) node{$\mathbf{\epsilon}$};
\draw (0.4,-1) node{$\mathbf{-\epsilon}$};
\draw[gray, thick] (0,0) -- (0,4);
\draw[gray, thick] (0,0) -- (0,-4);
\draw[gray, thick] (0,0) -- (4,0);
\draw[gray, thick] (0,0) -- (-4,0);
\draw
[
postaction={decorate,decoration={markings , 
mark=at position 0.20 with {\arrow[blue,line width=0.5mm]{>};}}}
][blue, thick] (-4,0.5)--(0,0.5);
\draw[blue, thick] (0,0.5)--(4,0.5);
\draw[gray, thick] (2,3) -- (2,3.4);
\draw[gray, thick] (2,3) -- (2.5,3);
\draw (2.3,3.3) node{$\mathbf{u}$};
\end{tikzpicture}
\caption{Contour $\rm{C_{even}}$ in the $u$-plane for even $AdS_{b}$} \label{fig4}
\end{figure}

 \begin{figure}[h]
\centering
\begin{tikzpicture}[thick,scale=0.85]
\filldraw[magenta] 
                (0,1) circle[radius=3pt]
                (0,-1) circle[radius=3pt];
\draw [decorate,decoration=snake] (0,-1) -- (0,-4);
\draw [decorate,decoration=snake] (0,1) -- (0,4);
\draw
[
postaction={decorate,decoration={markings , 
mark=at position 0.20 with {\arrow[gray,line width=1mm]{>};}}}
]
[
postaction={decorate,decoration={markings , 
mark=at position 0.70 with {\arrow[gray,line width=1mm]{>};}}}
]
(0,0) circle[radius=0.8cm];
\draw [gray,thick](1,1) node{$\mathbf{\rm{C_{odd}}}$};
\draw (0.4,1) node{$\mathbf{\epsilon}$};
\draw (0.4,-1) node{$\mathbf{-\epsilon}$};
\draw[gray, thick] (0,0) -- (0,4);
\draw[gray, thick] (0,0) -- (0,-4);
\draw[gray, thick] (0,0) -- (4,0);
\draw[gray, thick] (0,0) -- (-4,0);
\draw[gray, thick] (2,3) -- (2,3.4);
\draw[gray, thick] (2,3) -- (2.5,3);
\draw (2.3,3.3) node{$\mathbf{u}$};
\end{tikzpicture}
\caption{ Contour $\rm{C_{odd}}$ in the $u$-plane for even $AdS_{d+1}$} \label{fig5}
\end{figure}
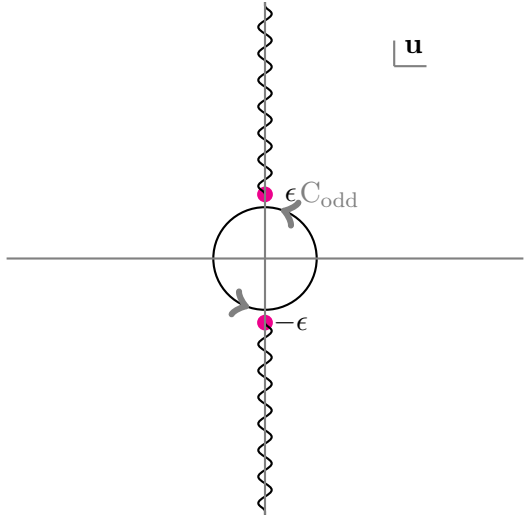
\end{subsection}

 \section{Conformal higher derivative Scalar}
 In the previous section, we have demonstrated that the partition functions of conformally coupled scalar fields are identical on conformally related spaces. In this section, we study the partition function of the conformal higher derivative fields and evaluate the partition function in terms of character integrals. We factorize the kinetic operator in terms of the two-derivative operators. Therefore the free energy becomes a simple addition of free energies of all the two-derivative operators. One can understand the character of the higher derivative fields as an addition of the character of two-derivative operators with a particular mass coming from the curvature coupling. We obtain the entanglement entropy and the conformal dimension of the twist operator of Weyl invariant higher derivative fields across a sphere from the partition function of the hyperbolic cylinder.
 \subsection{\texorpdfstring{$\partial^4$}{TEXT} operator}
 The Weyl invariant four-derivative scalar operator in $d>4$ dimension is given by \cite{Beccaria:2017dmw}
 \begin{align}
 \begin{split}\label{fourderoperator}
     \hat{\mathcal{O}}^{(4)}_{S^d}&=\hat{D}^4+\frac{4}{d-2}R^{\mu\nu}\hat{D}_{\mu}\hat{D}_{\nu}+k_d R \hat{D}^2+\frac{d-4}{2}\left(n_d R_{\mu\nu}R^{\mu\nu}+m_d R^2\right)+\mathcal{O}(\hat{D}^2 R),\\
     k_d&=-\frac{d^2-4 d+8}{2 (d-1) (d-2)},\qquad n_d=-\frac{2}{(d-2)^2},\qquad m_d=\frac{d^3-4 d^2+16 d-16}{8 (d-1)^2 (d-2)^2}.
 \end{split}
 \end{align}
 \subsection*{\texorpdfstring{$\partial^4$}{TEXT} operator on \texorpdfstring{$S^d_q$}{TEXT}} We evaluate the partition function of Weyl invariant four-derivative scalar field on branched sphere $S^d_q$. 
 The metric on the branched sphere is given by 
 \begin{equation}
 ds^2|_{S_q^{d}}  = \cos^2 \phi d \tau^2 + d\phi^2 + \sin^2 \phi d\Omega_{d-2}^2 ,
 \end{equation}
 where $\tau \sim \tau +2\pi q$ and $ 0\leq \phi \leq \frac{\pi}{2}$.
Since we are interested  to evaluate the partition function on homogenous space , we neglect the $\mathcal{O}(D^2 R)$ term in \eqref{fourderoperator}. The Ricci tensor and the curvature on this space is given by
\begin{eqnarray}
R_{\mu\nu}=\frac{R}{d}g_{\mu\nu},\qquad R=d(d-1).
\end{eqnarray}
We substitute the values of the parameters in \eqref{fourderoperator} and factorise it in two-derivative operators 
\begin{align}\label{factorisationfour}
     \hat{\mathcal{O}}^{(4)}_{S^d}&=\left(-\Delta_0+\frac{d^2-2d-8}{4}\right)\left(-\Delta_0+\frac{d^2-2d}{4}\right).
\end{align}
$\Delta_0$ is the spin-0  Laplacian on $S^d$;$\hspace{3 mm}$ $\Delta_0=\hat{D}^{\mu}\hat{D}_{\mu}$ . The eigenvalue and their corresponding  degeneracies  for  the scalar Laplacian on  the branched sphere
  are labelled by $2$ integers \cite{DeNardo:1996kp}.
 \begin{equation}\label{eigenvalue}
     \lambda^{(0)} _{n,m} =(n+\frac{m}{q})(n+\frac{m}{q}+d-1), \qquad  n, m \in \{ 0, \cdots \infty \} 
 \end{equation}
 with degeneracies 
 \begin{eqnarray} \label{degeneracy}
         g^{(0)}_{n,m=0}&=&\binom{n+d-2}{d-2},    \qquad n  \in \{ 0, \cdots \infty \}  \\  \nonumber
         g^{(0)}_{n,m>0}&=&2\binom{n+d-2}{d-2}, \qquad  n  \in \{ 0, \cdots \infty \}, \qquad m \in \{ 1, \cdots \infty\} 
 \end{eqnarray}
Therefore free energy can be written as sum of the free energies of two $2$-derivative operators. Note that the factorization of the higher derivative operators as a product of two-derivative operators can be done even if the two-derivative operators have the zero-modes in the spectrum. Therefore we can write the free energy as
\begin{align}
  \mathcal{F}^{(4)}_{S^d_q}&=\frac{1}{2}\rm{Tr}\log\mathcal{\hat{O}}^{(4)}_{S^d_q}\nonumber\\
  & =-\int_0^\infty \frac{d\tau}{2\tau}  \sum_{n,m =0}^\infty
    g_{n,m }^{(0)}\left( (  e^{ - \tau ( \lambda_{n,m}^{(0)}+\frac{d^2-2d-8}{4} )}  - e^{-\tau} )+(  e^{ - \tau ( \lambda_{n,m}^{(0)}+\frac{d^2-2d}{4} )}  - e^{-\tau} )\right) .
\end{align}
  The sum of the degeneracy vanishes due to the similar reason given in  \eqref{Hubbard} and we proceed with the first term to arrive
\begin{align}
   - \mathcal{F}^{(4)}_{S^d_q}&=\int_0^\infty \frac{d\tau}{2\tau}  \sum_{n,m =0}^\infty g_{n,m }^{(0)}\left(   e^{ - \tau ( (n+\frac{m}{q}+\frac{d-1}{2})^2+\frac{9}{4})} + e^{ - \tau ( (n+\frac{m}{q}+\frac{d-1}{2})^2+\frac{1}{4})}  \right).
\end{align}
Following the steps  from \eqref{contint} to \eqref{u part} we obtain
\begin{align}
  - \mathcal{F}^{(4)}_{S^d_q}  =
\int_{C}\frac{du}{ 2\sqrt{u^2+\epsilon^2}}\left(e^{-\nu_1\sqrt{u^2+\epsilon^2}}f^{(4)}_{0,q}    (u)+e^{-\nu_2\sqrt{u^2+\epsilon^2}}f^{(4)}_{0,q}(u)\right).
\end{align}
with $i\nu_1=\frac{3}{2}$; $i\nu_2=\frac{1}{2}$ and 
\begin{align}
    f^{(4)}_{0,q}(u)=\sum_{n,m=0}^{\infty}g_{n,m }^{(0)}e^{i u(n+\frac{m}{q}+\frac{d-1}{2})}
\end{align}
We now deform the contour 
$C$ from the real line to the contour  $C'$ which runs on the both sides of the branch cut on the imaginary axis 
originating at $u = i \epsilon$ on the  $u$-plane.
Substituting $u = i t$ we obtain
\begin{equation} 
 - \mathcal{F}^{(4)}_{S^d}  =
\int_{C'}\frac{dt}{ 2\sqrt{t^2-\epsilon^2}}\left(\left(e^{i\nu_1\sqrt{t^2-\epsilon^2}}+e^{-i\nu_1\sqrt{t^2-\epsilon^2}}\right)
+\left(e^{i\nu_2\sqrt{t^2-\epsilon^2}}+e^{i\nu_2\sqrt{t^2-\epsilon^2}}\right)f^{(4)}_{0,q}(u)\right).
\end{equation}
We can now take $\epsilon\rightarrow 0$ and obtain
\begin{align}\label{fourdersphere}
-\mathcal{F}^{(4)}_{S^d_q} &=\int_0^{\infty}\frac{dt}{2t}\frac{1+e^{-\frac{t}{q}}}{1-e^{-\frac{t}{q}}}\frac{e^{-\frac{1}{2} (d-2) t}+e^{-\frac{d t}{2}}+e^{-\frac{1}{2} (d-4) t}+e^{-\frac{1}{2} (d+2) t}}{\left(1-e^{-t}\right)^{d-1}} \nonumber\\
 &=\int_0^{\infty}\frac{dt}{2t}\frac{1+e^{-\frac{t}{q}}}{1-e^{-\frac{t}{q}}}\left(\ \chi_{ (d, 0)\;{\nu_1 }  }^{dS} ( t)+ \chi_{ (d, 0)\;{\nu_2 }  }^{dS} ( t)\right).
\end{align}
$ \chi_{ (d, 0)\;{\nu_1 }  }^{dS} ( t)$ and $\chi_{ (d, 0)\;{\nu_2 }  }^{dS} ( t)$ corresponds to the  characters of the two factorised operators and the parameters $\nu_1$ and $\nu_2$ are determined from the masses of each two-derivative operators from the relation given by
  \begin{eqnarray}\label{definu}
  i \nu =  \sqrt{  \frac{(d-1)^2}{4}  -   m^2}.
  \end{eqnarray}
  Then character representation of massive scalar field of  de Sitter group $SO(1, d)$ is given by   \cite{Anninos:2020hfj}  
\begin{equation}\label{chi}
\chi_{(d, 0) \nu}^{dS} (t) = \frac{ e^{ - t\Delta_+} + e^{- t\Delta_-} }{ ( 1- e^{-t} )^{d-1}}, 
\qquad \qquad \Delta_{\pm} = \frac{d-1}{2} \pm i \nu .
\end{equation}
Note that $\Delta_{\pm}$ for the two operators are
\begin{align*}\label{defcharacter}
    \Delta^{(1)}_{\pm}=\frac{d-1}{2} \pm \frac{3}{2},\qquad \Delta^{(2)}_{\pm}=\frac{d-1}{2} \pm \frac{1}{2}.
\end{align*}
  Therefore the character of the $4$-derivative Weyl-invariant scalar operator becomes the addition of the characters of two $2$-derivative spin-0 massive scalar operators. 
  \subsection*{\texorpdfstring{$\partial^4$}{TEXT} operator on \texorpdfstring{$AdS_{d-1}\times S^1_q$}{TEXT}} We have evaluated the partition function of the Weyl invariant $4$-derivative operator on $S^d_{~q}$. Since the theory is Weyl invariant one can expect the partition function should remain the same on the conformally related spaces.  It is well known that the hyperbolic cylinder and the branched sphere can be related by the Weyl transformation. Therefore we express the partition function on the hyperbolic cylinder in terms of integral over characters and verify whether it agrees with the branched sphere.
  
  The Ricci tensor and the curvature on $S^1_q\times AdS_{d-1}$ is given by
  \begin{eqnarray}
  R_{00}=0,\qquad R_{0i}=0,\qquad R_{ij}=\frac{R}{d-1}g_{ij},\qquad R=-(d-1)(d-2).
  \end{eqnarray}
  Here $\{i,j\}$ denote the directions along $AdS_{d-1}$.
  Substituting these values in \eqref{fourderoperator} and factorising it into two quadratic operators we get \cite{Beccaria:2017dmw}
  \begin{align}
   \mathcal{\hat{O}}^{(4)}_{S^1_q\times AdS_{d-1} }&=   \bigg[-\partial_0^2+\bigg\{\sqrt{-\Delta^{\rm{AdS}}_0-\frac{(d-2)^2}{4}}+i\bigg\}^2\bigg]\bigg[-\partial_0^2+\bigg\{\sqrt{(-\Delta^{\rm{AdS}}_0-\frac{(d-2)^2}{4})}-i\bigg\}^2\bigg].
  \end{align}
  $\Delta^{\rm{AdS}}_0$ is the spin-0 zero Laplacian on $AdS_{d-1}$.  The eigen values are given in \eqref{eigen}. Note that the shift in the eigen value cancels the factor $\frac{(d-2)^2}{4}$ inside the square root.
Using the eigen values  and the Kaluza-Klein decomposition of the partition function, the free energy becomes
\begin{align}\label{feq1}
\mathcal{F}^{(4)}_{S^1_q\times AdS_{d-1}}
&   =\frac{1}{2}\rm{Tr}\log\mathcal{\hat{O}}^{(4)}_{S^1\times AdS_{d-1}}\nonumber\\
&= \frac{1}{4}  \sum_{n=-\infty}^\infty \int_{-\infty}^\infty d\lambda \mu_{0}^{(d-1)}(\lambda) 
   \left( \log ( \frac{n^2}{q^2} +(\lambda+i)^2 )+\log ( \frac{n^2}{q^2} +(\lambda-i)^2
 )\right).
\end{align}
 The free energy becomes addition  of free energies of two factorised quadratic operators.
 We can rewrite the logarithm using the identity given in (\ref{logiden})  to obtain
 \begin{align}
   -\mathcal{F}^{(4)}_{S^1_q\times AdS_{d-1}}&=\frac{1}{4} \int_0^\infty \frac{d\tau}{\tau} \sum_{n=-\infty}^\infty 
  \int_{-\infty}^{\infty} d\lambda 
  \mu_{0}^{(d-1)}(\lambda)\bigg[ ( e^{ -\tau( (\lambda+i)^2 + \frac{n^2}{q^2} )} - e^{-\tau} )   \nonumber\\
  &+( e^{ -\tau( (\lambda-i)^2 + \frac{n^2}{q^2} )} - e^{-\tau} )\bigg].
 \end{align}
We need to evaluate the integrals over Plancherel measure in order to obtain the partition function. Since Plancherel measure is an even function in $\lambda$ one can extend the lower limit of the integral all the way to $-\infty$. Therefore one can write
\begin{align*}
    \int_{-\infty}^{\infty}\mu_0^{(d-1)}(\lambda)d\lambda=\lim_{u\rightarrow 0}W_0^{(d-1)}(u),
\end{align*}
where $W_0^{(d-1)}(u)$ is the Fourier transform of the Plancherel measure of spin-0 Laplacian on $AdS_{d-1}$. The expression of $W_0^{(d-1)}( u)$ can be obtained by substituting $b=(d-1)$ in \eqref{PlFourier}. From the expression of $W_0^{(d-1)}( u)$  one can observe  that $\lim_{u\rightarrow 0}W_0^{(d-1)}(u)$ vanishes for sufficiently large negative $d$. Therefore the integral of the Plancherel measure also vanishes.
\begin{equation}
\int_0^\infty d\lambda \mu_0^{(d-1)} =0 .
\end{equation}
So we proceed  with the first term.
\begin{align}
 -\mathcal{F}^{(4)}_{S^1_q\times AdS_{d-1}}&
  =
  \int_0^{\infty}\frac{d\tau}{4\tau}e^{-\frac{\epsilon^2}{4\tau}}\int_{-\infty}^{\infty} 
  d\lambda 
  \mu^{(d-1)}_0 (\lambda) \big[ \left(  e^{ - \tau (\lambda+i)^2} + 2 \sum_{n=1}^\infty 
  e^{-\tau(\lambda^2+\frac{n^2}{q^2})  }\right)\nonumber\\
  &+.\left(  e^{ - \tau (\lambda-i)^2} + 2 \sum_{n=1}^\infty 
  e^{-\tau(\lambda^2+\frac{n^2}{q^2})  }\right)\big].
\end{align}
We perform the  integral over $\lambda$ using Hubbard-Stratonovich trick which linearises the sum over Kaluza-Klein modes along $S^1_q$ direction
\begin{align}\label{lastustep}
    -\mathcal{F}^{(4)}_{S^1_q\times AdS_{d-1}}&=\frac{1}{4}\int_{-\infty}^{\infty}du\int_0^{\infty}\frac{d\tau}{\sqrt{4\pi\tau^3}} \sum_{n=0}^{\infty}e^{-\frac{\epsilon^2+u^2}{4\tau}}e^{-\tau \frac{n^2}{q^2}}\nonumber\\
   & \times\int_{-\infty}^{\infty}d\lambda  \mu^{(d-1)}_0 (\lambda)(e^{i(\lambda+i)u}+e^{i(\lambda-i)u})\nonumber\\
   &=\frac{1}{4}\int_Cdu\int_0^{\infty}\frac{d\tau}{\sqrt{4\pi\tau^3}} \sum_{n=0}^{\infty}e^{-\frac{\epsilon^2+u^2}{4\tau}}e^{-\tau\frac{n^2}{q^2}}W_0^{(d-1)}( u)(e^u+e^{-u}).
\end{align}
$W_0^{(d-1)}( u)$ is the Fourier transform of the Plancherel measure which is defined earlier.
Following the steps in \eqref{KKsum} and \eqref{uintegralads} in even $AdS_{d-1}$   we obtain
\begin{align}\label{fourderhyp}
 -\mathcal{F}^{(4)}_{S^1_q\times AdS_{d-1}}& =\int_0^{\infty}\frac{du}{2u} \frac{1+e^{-\frac{u}{q}}}{1-e^{-\frac{u}{q}}}\frac{e^{-\frac{1}{2} (d-2) u}}{\left(1-e^{-u}\right)^{d-2}} \frac{1+e^{-u}}{1-e^{-u}}(e^{-u}+e^u)\nonumber\\
 &=\int_0^{\infty}\frac{du}{2u} \frac{1+e^{-\frac{u}{q}}}{1-e^{-\frac{u}{q}}}\frac{e^{-\frac{1}{2} (d-2) u}+e^{-\frac{d u}{2}}+e^{-\frac{1}{2} (d-4) u}+e^{-\frac{1}{2} (d+2) u}}{(1-e^{-u})^{d-1}}.
\end{align}
Similarly we proceed in odd $AdS_{d-1}$ dimension where the contour is given in figure [\ref{fig4}]. The integral is evaluated by taking the residue at $u=0$.
\begin{align}\label{Oddfour}
  -\mathcal{F}^{(4)}_{S^1_q\times AdS_{d-1}}&   =\frac{\log R}{2\pi i}\int_{\rm{C_{Odd}}}\frac{du}{2u} \frac{1+e^{-\frac{u}{q}}}{1-e^{-\frac{u}{q}}}\frac{e^{-\frac{1}{2} (d-2) u}+e^{-\frac{d u}{2}}+e^{-\frac{1}{2} (d-4) u}+e^{-\frac{1}{2} (d+2) u}}{(1-e^{-u})^{d-1}}.
\end{align}
Here $R$ is the ratio of the radial cutoff on $AdS$ to the radius of $AdS$.\\
Note that the expression of the partition functions given in \eqref{fourderhyp} and \eqref{fourdersphere} are identical. The partition functions are expressed as the integral transform of the character. Therefore the equality of the partition functions becomes obvious. We are also interested in evaluating the entanglement entropy of the four-derivative Weyl-covariant scalar field across a spherical entangling surface. Given the partition function of a conformal field on a hyperbolic cylinder, one can evaluate the universal contribution of  R\'{e}nyi entropy from the logarithmic divergence of the free energy in an even dimension. Therefore we expand the integrand \eqref{fourdersphere}  around $t=0$ and collect the coefficient of $1/t$ term. This is the same as evaluating the integral \eqref{Oddfour} by taking the residue at $u=0$. The R\'{e}nyi entropy $S_q$ and the entanglement entropy  $S_{\rm{EE}}$ are given by 
     \begin{eqnarray}
          S_q=\frac{- \mathcal{F}^{(4)}_q+q \mathcal{F}^{(4)}_{q=1}}{1-q}, \qquad\qquad
          S_{\rm{EE}}=\lim_{q\rightarrow 1}S_q.
    \end{eqnarray}
      In table \ref{table1}  we have listed both these entropies for the four-derivative Weyl-covariant scalar 
    for even $4 \leq d \leq 14$. In $d=4$ the R\'{e}nyi entropy $S_q$ \footnote{We have used position space cutoff $\epsilon\sim\frac{1}{\Lambda_{UV}}$ and therefore we get an extra minus sign.} and the entanglement entropy  $S_{\rm{EE}}$ agree with \cite{Beccaria:2014jxa}.
    
    \begin{table}[ht]
\centering { \footnotesize{
\begin{tabular}{c|l|c}
\hline
$d $ & \quad $~~~~~~~~~~~~~~~~~~~~~~~S_q $ & $S_{\rm{EE}}$ \\
\hline 
& &  \\
$4$ & \quad$-\frac{(q+1) \left(29 q^2-1\right)}{180 q^3} $ & $-\frac{14}{45}$
   \\   
&  & \\
$6$ & \quad$ \frac{(q+1) \left(33 \left(q^4+q^2\right)-2\right)}{15120 q^5}$ & $ \frac{8}{945}$  \\
& &  \\
$8$ & \quad$-\frac{(q+1) \left(q^2 \left(191 \left(q^4+q^2\right)+37\right)-3\right)}{907200 q^7} $ &  $-\frac{13}{14175}$  \\
&  & \\
$10 $& \quad$\frac{(q+1) \left(q^2 \left(\left(3455 \left(q^4+q^2\right)+947\right) q^2+89\right)-10\right)}{119750400 q^9} $ &
$\frac{62}{467775}$  \\
& &  \\
$10$& \quad $-\frac{(q+1) \left(q^2 \left(\left(9 \left(338131 \left(q^4+q^2\right)+105899\right) q^2+160871\right) q^2+7718\right)-1382\right)}{653837184000 q^{11}}$
& $-\frac{28151}{1277025750}$\\
& & \\ 
$14 $& \quad 
$\frac{(q+1) \left(13 q^2 \left(\left(\left(\left(498053 \left(q^4+q^2\right)+167429\right) q^2+35253\right) q^2+4134\right) q^2+74\right)-420\right)}{7846046208000 q^{13}}$
&  $\frac{7636}{1915538625} $\\
& &\\
\hline
\end{tabular}
\caption{ Universal terms in R\'{e}nyi entropies and entanglement entropies of four-derivative Weyl-covariant scalar.  }
\label{table1}
}}
\end{table}
For conformal field theory on even dimensional flat space the first derivative of  R\'{e}nyi entropy $S_q$ at $q=1$ is related to the central charge of the theory $C_{T,d}$ by the following relation \cite{Perlmutter:2013gua}
\begin{align}\label{check}
    \partial_q S_q|_{q=1}=-\rm{Vol}(AdS_{d-1})\pi^{\frac{d}{2}+1}\frac{\Gamma[\frac{d}{2}](d-1)}{(d+1)!(\rm{Vol}(S^{d-1})^2}C_{T,d},
\end{align}
where $\rm{Vol}(S^{d-1})$ is the volume of sphere in $(d-1)$ dimension and $\rm{Vol}(AdS_{d-1})$ is the regularised volume of $AdS_{d-1}$ in odd dimension given in \eqref{volads}.
For $4$-derivative conformal scalar central charge is given by \cite{Osborn:2016bev}
\begin{align}
    C_{T,d}^{(4)}=-\frac{2 d (d+4)}{(d-1) (d-2)}.
\end{align}
We have verified that the  R\'{e}nyi entropies $S_q$ presented in table \eqref{table1} for $4\leq d\leq 14 $ satisfy the relation \eqref{check}.
We now evaluate the conformal dimension of the co-dimension two twist operator located at the entangling surface. 
\begin{table}[ht]
\centering { \footnotesize{
\begin{tabular}{c|c|c|c}
\hline
$d $ & $h_q$ & $ h_q'|_{q=1}$  & $ h_q^{\prime\prime}|_{q=1}$  \\
\hline 
& &  \\
$4$ & $-\frac{9 q^4-10 q^2+1}{360 \pi  q^3} $ & $-\frac{2}{45 \pi } $ & $\frac{1}{45 \pi } $
   \\   
&  &  & \\
$6$ &  $ -\frac{19 q^6-21 q^2+2}{15120 \pi ^2 q^5} $ & $ -\frac{1}{210 \pi ^2}$  &  $\frac{4}{315 \pi ^2}$ \\
& &  &  \\
$8$ & $-\frac{641 q^8-462 q^4-200 q^2+21}{2116800 \pi ^3 q^7} $ &  $-\frac{1}{735 \pi ^3} $  & $\frac{6}{1225 \pi ^3} $\\
&  &  & \\
$10 $& $-\frac{4139 q^{10}-2508 q^6-1430 q^4-231 q^2+30}{29937600 \pi ^4 q^9}$ & 
$-\frac{1}{1485 \pi ^4} $  & $\frac{289}{103950 \pi ^4} $  \\
& &  &  \\
$12$& $\frac{-11370133 q^{12}+6270264 q^8+3961100 q^6+1072071 q^4+81900 q^2-15202}{119870150400 \pi ^5 q^{11}}$
& $ -\frac{16}{33033 \pi ^5} $ & $\frac{7508}{3468465 \pi ^5}$ \\
& & & \\ 
$14 $& 
$\frac{-24802367 q^{14}+12894336 q^{10}+8591440 q^8+2831829 q^6+475020 q^4+15202 q^2-5460}{283329446400 \pi ^6 q^{13}}$
&  $-\frac{6}{13013 \pi ^6} $  & $\frac{593}{273273 \pi ^6} $\\
&  & & \\
\hline
\end{tabular}
\caption{Conformal dimension of the twist operator of the four-derivative Weyl invariant scalar. }
\label{table2}
}}
\end{table}
The first and second derivatives of the conformal dimension of the twist operator with respect to $q$ at $q=1$  can be related to the parameters determining the two and three-point functions of the stress tensor \cite{Hung:2014npa, Osborn:1993cr}. So we evaluate the first and second derivatives of $h_q$ with respect to $q$ at $q=1$ which will be useful to understand the causality or unitarity of the higher derivative fields which we will discuss in section \eqref{causality}.
\subsection{\texorpdfstring{$\partial^6$}{TEXT} operator }
In the previous section, we have shown that the partition function of the conformal four-derivative scalar is identical on the conformally equivalent classes of space. Therefore one expects that $6$-derivative conformal scalars should also have the same partition function on conformally equivalent spaces like $S^d$ and $AdS_{d-1}\times S^1$. In this section, we study six-derivative Weyl-invariant scalar field in $d=6$-dimensions and evaluate the entanglement entropy, conformal dimension of the twist operator across a spherical entangling surface.\\
The Weyl-invariant $6$-derivative scalar operator in  $d=6$ dimensions is given by \cite{Beccaria:2017dmw}
 \begin{align}\label{sixdimder}
 \begin{split}
     \hat{\mathcal{O}}^{(6)}_{S^6_q}&=-\hat{D}^6-(16 P_{\mu\nu}-6g_{\mu\nu}P)\hat{D}^{\mu}\hat{D}^{\nu}\hat{D}^2\nonumber\\
     &+8(4P^{\mu\nu}P-g^{\mu\nu}P_{\alpha\rho}P^{\alpha\rho})\hat{D}_{\mu}\hat{D}_{\nu}+8(P_{\mu\nu}P^{\mu\nu}-P^2)\hat{D}^2.\\
 \end{split}
 \end{align}
The expression of Schouten tensor $P_{\mu\nu}$ and its trace $P$ in arbitrary dimension are known \cite{Osborn:2015rna}
\begin{align}
   P_{\mu\nu}&=\frac{1}{d-2}( R_{\mu\nu}-\frac{1}{2(d-1)}g_{\mu\nu}R),\qquad P=P_{\mu}^{\mu}.
\end{align}
\subsection*{\texorpdfstring{$\partial^6$}{TEXT} operator on \texorpdfstring{$S_q^6$}{TEXT}}
We start by evaluating the partition function of Wel-invariant scalar on branched sphere $S^6_q$. The Ricci tensor and the curvature on sphere in 6 dimension is given by  $R_{\mu\nu}=5g_{\mu\nu}$ and $R=30$.\\
Substituting $P_{\mu\nu}$ and $P$ in \eqref{sixdimder} and factorising it we obtain
\begin{align}
   \hat{\mathcal{O}}^{(6)}_{S^d}=-\Delta_0^{S^6}(-\Delta_0^{S^6}+4)(-\Delta_0^{S^6}+6),\nonumber\\ 
\end{align}
where $\Delta_0^{S^6}=\hat{D}^{\mu}\hat{D}_{\mu}$ is the spin-0 Laplacian on $S^6$. Note that the factorization makes sense even if we consider the zero-modes.
 The eigenvalue and their corresponding  degeneracy  for  the scalar Laplacian on  the branched sphere
  are labelled by $2$ integers
 \begin{equation}\label{eigenvaluebranched}
     \lambda^{(0)} _{n,m} =(n+\frac{m}{q})(n+\frac{m}{q}+5), \qquad  n, m \in \{ 0, \cdots \infty \} 
 \end{equation}
 with degeneracy 
 \begin{eqnarray} \label{degeneracybranched}
         g^{(0)}_{n,m=0}&=&\binom{n+4}{4},    \qquad n  \in \{ 0, \cdots \infty \}  \\  \nonumber
         g^{(0)}_{n,m>0}&=&2\binom{n+4}{4}, \qquad  n  \in \{ 0, \cdots \infty \}, \qquad m \in \{ 1, \cdots \infty\} 
 \end{eqnarray}

Therefore the free energy can be written as 
\begin{align}
    \mathcal{F}^{(6)}_{S^6_q}&=\frac{1}{2}\rm{Tr}\log\mathcal{\hat{O}}^{(6)}_{S^6_q}\nonumber\\
  & =-\int_0^\infty \frac{d\tau}{2\tau}  \sum_{n,m =0}^\infty
    g_{n,m }^{(0)}\left( (  e^{ - \tau  \lambda_{n}^{(0)}}  - e^{-\tau} )+(  e^{ - \tau ( \lambda_{n}^{(0)}+4 )}  - e^{-\tau} )+(  e^{ - \tau ( \lambda_{n}^{(0)}+6 )}  - e^{-\tau} )\right) .
\end{align}
Following the steps given from \eqref{contint} to \eqref{u part} we obtain
 \begin{eqnarray} \label{sphere6}
     - \mathcal{F}^{(6)}_{S^6_q}  = \int_{\epsilon}^{\infty}\frac{dt}{2\sqrt{t^2-\epsilon^2}}\left(\sum_{i=1}^3e^{i\nu_i\sqrt{t^2-\epsilon^2}}+e^{-i\nu_i\sqrt{t^2-\epsilon^2}}\right)f_{q,6}^{(0)} (i t).\
    \end{eqnarray}
    with $i\nu_1 = \frac{5}{2},\qquad i\nu_2 = \frac{3}{2},\qquad i\nu_3 = \frac{1}{2} $ and 
    \begin{eqnarray}\label{fq0}
    f_{q,6}^{(0)} ( u ) &=& \sum_{n,m =0}^\infty g^{(0)}_{n,,m}e^{i(n+\frac{m}{q}+\frac{5}{2})u}\nonumber\\
    &=& \frac{e^{ \frac{i u 5}{2}}}{ ( 1 - e^{i u } )^5 } \frac{ 1 + e^{ i \frac{ u }{ q}} }{ 1 - e^{ i \frac{ u }{ q}}}.
    \end{eqnarray}
    We now substitute $ f_{q,6}^{(0)} ( u )$ in \eqref{sphere6} and take $\epsilon\rightarrow 0$ to obtain
    \begin{align}\label{finalsixdersphere}
 -  \mathcal{F}^{(6)}_{S^6_q} &=\int_0^{\infty}\frac{dt}{2t}\frac{1+e^{-\frac{t}{q}}}{1-e^{-\frac{t}{q}}}\frac{(1+e^{-5t})+(e^{-4t}+e^{- t})+(e^{-2 t}+e^{-3t})}{\left(1-e^{-t}\right)^5} \nonumber\\
 &=\int_0^{\infty}\frac{dt}{2t}\frac{1+e^{-\frac{t}{q}}}{1-e^{-\frac{t}{q}}}\left(\ \chi_{ (5, 0)\;{\nu_1 }  }^{dS} ( t)+ \chi_{ (5, 0)\;{\nu_2 }  }^{dS} ( t)+\chi_{ (5, 0)\;{\nu_3 }  }^{dS} ( t)\right).
\end{align}
The expressions of $\chi_{ (5, 0)\;{\nu_i }  }^{dS} ( t)$ can be obtained from \eqref{chi} by substituting the values of $\nu_i$ for each two-derivative operators in $d=6$ dimension.
Note that the partition function of the six-derivative conformal scalar on $S^6_q$ can be expressed as an integral transform of the characters of the factorized 2-derivative operators. Therefore the character of higher derivative Weyl-invariant scalar can be written as a sum of the characters of the 2-derivative massive scalar operators where the mass of each 2-derivative operator comes from the curvature coupling.
\subsection*{\texorpdfstring{$\partial^6$}{TEXT} operator on \texorpdfstring{$AdS_5\times S^1_q$}{TEXT}} The hyperbolic cylinder $AdS_5\times S^1_q$ is conformally related to $S^6_q$.
The curvature induced mass comes only from the $AdS_5$.
 The Ricci tensor and the curvature are given by
\begin{eqnarray}
R_{0i}=0,\qquad R_{ij}=-4g_{ij},\qquad R=-20.
\end{eqnarray}
Here $\{i,j\}$ denotes the direction along the $AdS_5$.
We substitute the curvature in the expression of Schouten tensor and factorise the operator in terms of three two-derivative operators \cite{Beccaria:2017dmw} to obtain
\begin{align}
   \hat{\mathcal{O}}^{(6)}_{AdS_5\times S^1_q}=(-\partial_{\tau}^2-\Delta_0^{AdS_5}-1)(-\partial_{\tau}^2-(\sqrt{-\Delta_0^{AdS_5}-1}+2i)^2)(-\partial_{\tau}^2-(\sqrt{-\Delta_0^{AdS_5}-1}-2i)^2).\nonumber\\ 
\end{align}
Therefore we get a combination of a 2-derivative conformal scalar and two conjugate operators with the shift in the masses.
$\Delta^{\rm{AdS_5}}_0$ is the spin-0 zero Laplacian on $AdS_5$
  \begin{equation}
 \Delta^{\rm{AdS_5}}_0\psi^{\{\lambda, u\}} = -\left[  \lambda^2 + 1 \right] 
 \psi_\lambda^{\{\lambda, u\}} ,
 \end{equation}
  $\psi^{ \{\lambda, u\} }$  are the corresponding eigen functions, $\{u\}$ labels other quantum numbers
 on $AdS_5$. We follow the steps from \eqref{feq1} to \eqref{lastustep} and obtain
 \begin{align}
 -\mathcal{F}^{(6)}_{AdS_5\times S^1_q} & = \frac{\log R}{4\pi i}\int_{\rm{C_{Odd}}}du\int_{0}^{\infty}\frac{{d\tau}}{{\sqrt{4\pi\tau^3}}} \sum_{n=0}^{\infty}e^{-\frac{\epsilon^2+u^2}{4\tau}}e^{-\tau\frac{(n)^2}{q^2}}W_0^{(5)}( u)(1+e^{2u}+e^{-2u}).
 \end{align}
 $W_0^{(5)}$ is the Fourier transform of the Plancherel measure of the spin-0 field on $AdS_5$. The expression of $W_0^{(5)}( u)$ can be obtained by substituting $b=5$ in \eqref{PlFourier}. Now we perform the  integral over $\tau$ and sum over the Kaluza-Klein modes along $S^1_q$ to obtain
 \begin{align}\label{finalsixads}
      -\mathcal{F}^{(6)}_{AdS_5\times S^1_q} & =\frac{\log R}{2\pi i} \int_{\rm{C_{Odd}}}\frac{du}{2u}\frac{1+e^{-\frac{u}{q}}}{1-e^{-\frac{u}{q}}}\frac{(1+e^{-5u})+(e^{-4u}+e^{- u})+(e^{-2 u}+e^{-3u})}{\left(1-e^{-u}\right)^5}.
 \end{align}
The integrand has the same form as it is given in \eqref{finalsixdersphere}. Note that we have done the sum over all the Kaluza-Klein modes including the zero-mode. The integrand organizes itself in a nice character form when we include the zero-mode. In the literature usually, this is omitted which results in the absence of a term linear in $q$ in the logarithmic part of the free energy. However, the R\'{e}nyi entropy and the entanglement entropy do not change under the shift of the free energy. Therefore we expand the integrand \eqref{finalsixdersphere} around $t=0$  and collect the co-efficient of $1/t$ term to obtain the log-divergent piece of the free energy. This is the same as the evaluation of the residue of the integral \eqref{finalsixads} at $u=0$. The universal part of the R\'{e}nyi entropy and entanglement entropy can be obtained from the free energy 
 \begin{align}
          S_q&=-
          \frac{(q+1) \left(1577 q^4-103 q^2+2\right)}{10080 q^5}, \qquad\qquad
          S_{\rm{EE}}=-\frac{41}{140}.
    \end{align}
    The  R\'{e}nyi entropy and entanglement entropy agree with \cite{Beccaria:2017dmw} in $d=6$. The first derivative of R\'{e}nyi entropy with respect to $q$ at $q=1$ is related to the central charge of the theory and the relation is given in \eqref{check}. The central charge for the $6$-derivative conformal scalar in $d=6$ is given by $ C_{T,d=6}^{(6)}=54$ \cite{Beccaria:2017dmw} and we verify that the relation \eqref{check} is satisfied.
   
    We obtain the weight of the twist operator across the entangling surface
    \begin{align}
        h_q&=\frac{q}{5 \rm{Vol}(AdS_5)}\left(\partial_q\mathcal{F}_q|_{q=1}-\partial_q\mathcal{F}_q\right)\nonumber\\
        &=\frac{275 q^6-336 q^4+63 q^2-2}{10080 \pi ^2 q^5}.
    \end{align}
    First and second derivative of weight of  twist operator $h_q$ with respect to $q$ at $q=1$ are related to the parameters $a$, $b$ and $c$ which characterize the two and three point function of the stress tensor. For $6$-derivative conformal scalar we obtain
    \begin{align}\label{6hder}
        h_q^{\prime}|_{q=1}=\frac{3}{70 \pi ^2}, \qquad h_q^{\prime\prime}|_{q=1}=\frac{1}{420 \pi ^2}.
    \end{align}
    These numbers will be useful to understand the causality of the $6$-derivative conformal scalar in $d=6$ dimension.
 \section{Weyl invariant fermionic fields}
 In the previous section, we have shown that the partition functions of higher derivative conformal scalars are identical on conformally equivalent spaces. In this section, we study the Weyl invariant fermionic fields on conformally related spaces. We express the partition function as the integral transform of the character. The character integral representation helps to understand the relation between partitions functions explicitly.
 \subsection{\texorpdfstring{$\slashed{\partial}$}{TEXT} operator}The action of Weyl-invariant massless spin-half field in $d$ dimension is given by
 \begin{align}
     S&=i\int\sqrt{g}d^dx \hspace{2mm}\Bar{\psi}\slashed{\nabla}\psi,
 \end{align}
 where $\slashed{\nabla}=\gamma^{\mu}\nabla_{\mu}$. One obtains the free energy of the massless fermionic field by using the spectrum of the squared operator $\slashed{\nabla}^2=\Delta_0-\frac{R}{4}$. Here $\Delta_0$ is the spin-0 Laplacian and $R$ is the curvature of the background.\\
\subsection*{\texorpdfstring{$\slashed{\partial}$}{TEXT} operator on \texorpdfstring{$S^d$}{TEXT}} Therefore the free energy becomes 
 \begin{align}\label{freespinhalf}
     \mathcal{F}_{\frac{1}{2}}&=-\rm{Tr}\log(i\slashed{\nabla})\nonumber\\
     &= - \frac{1}{2} 
    \sum_{n=0}^\infty  g_{n,d}^{(\frac{1}{2})} \log \left( \lambda_{n,d}^{(\frac{1}{2})}   \right) \nonumber\\
     &=\int_{\epsilon}^{\infty}\frac{d\tau}{2\tau}e^{-\frac{\epsilon^2}{4\tau}}\sum_{n=0}^{\infty}g^{(\frac{1}{2})}_{n,d} \left(e^{-\tau\lambda^{(\frac{1}{2})}_{n,d}}-e^{-\tau}\right).
 \end{align}
 In the last step we used the identity \eqref{logiden} to write the the logarithm of the eigen-value. To evaluate the partition function we require
 the eigen value and the degeneracy of the Dirac fermion on $S^d$ which are known \cite{Camporesi:1995fb}
 \begin{align}
 \begin{split}
      g^{(\frac{1}{2})}_{n,d}&=\frac{2^{\frac{d}{2}} (d+n-1)!}{(d-1)! n!},\\
     \lambda^{(\frac{1}{2})}_{n,d}&=(n+\frac{d}{2})^2.
 \end{split}
 \end{align}
 To evaluate the  partition function we perform  sum over degeneracy by using zeta-regularisation prescription and obtain
 \begin{align}\label{spinhalfsum}
     \sum_{n=0}^{\infty} g^{(\frac{1}{2})}_{n,d}=0.
 \end{align}
 We therefore drop the second term in \eqref{freespinhalf} and use Hubbard-Stratonovich trick to linearise the sum in the first term
 \begin{align}\label{spinhalfstepinitial}
     \mathcal{F}_{\frac{1}{2}}& =  - \int_{\epsilon}^\infty \frac{d\tau}{2\tau} e^{-\frac{\epsilon^2}{4\tau} } \int_{C} \frac{du}{\sqrt{4\pi\tau}}e^{-\frac{u^2}{4\tau}}f_{\frac{1}{2}}(u).
 \end{align}
Here the contour $C$ runs from $-\infty $ to $\infty$ slightly above the real axis as shown in figure [\ref{fig1}].
\begin{align}
   f_{\frac{1}{2}}(u) &=\sum_{n=0}^{\infty}g_{n,d}^{(\frac{1}{2})} e^{i u(n+\frac{d}{2})}\nonumber\\
&=\frac{2^de^{iu\frac{d}{2}}}{(1-e^{iu})^d}.
\end{align}
We now perform the $\tau$ integral and deform the contour 
$C$ from the real line to the contour  $C'$ which runs on the both sides of the branch cut on the imaginary axis 
originating at $u = i \epsilon$ on the  $u$-plane. Finally we substitute $u=it$ and obtain
\begin{align}\label{spinhalfspherepartition}
    \mathcal{F}_{\frac{1}{2}}&=-\int_0^{\infty}\frac{dt}{2t}\frac{2e^{-\frac{t}{2}}}{(1-e^{-t})}\frac{2^{d-1}e^{-t\frac{d-1}{2}}}{(1-e^{-t})^{d-1}}.
\end{align}
Therefore we obtain the free energy as integral transform of the Harish-Chandra character. Note that the kinetic factor $2\frac{e^{-\frac{t}{2}}}{1-e^{-t}}$ is different compared to the bosonic kinematic factor in the integral. Therefore
the $\rm{SO}(1.d)$ Harish-Chandra character of a massless Dirac fermion is given by
\begin{align}
   \chi_{ (d, \frac{1}{2})\;{\rm conf}  }^{dS} ( u)&=2^{d-2}\frac{e^{-t\frac{d-1}{2}}+e^{-t\frac{d-1}{2}}}{(1-e^{-t})^{d-1}}.
\end{align} This agrees with the equation (3.13) of \cite{Anninos:2020hfj} when we substitute $d=4$ and consider the massless case.
\subsection*{$\slashed{\partial}$ operator on $ S^a\times AdS_{b}$}
Since $S^a\times AdS_b$ is confromally related to $S^{a+b}$ one expects the partition function of conformally coupled spin-half field will be identical on these backgrounds.
Therefore we study the partition function of the massless Dirac Fermion on $S^a\times AdS_b$. The free energy of the Weyl-invariant massless fermionic field on $S^a\times AdS_b$ can be expressed as
\begin{align}
   \mathcal{F}^{\frac{1}{2}}_{S^a\times AdS_b}&= \frac{1}{4}\sum_{n=0}^{\infty}g^{(\frac{1}{2})}_n\int_{-\infty}^{\infty}\mu^{(\frac{1}{2})}_b(\lambda)\log\left(\lambda^2+(n+\frac{a}{2})^2\right).
\end{align}
Here we have used the eigen value of the spin-half Laplacian on $AdS_b$
\begin{align}
    \slashed{\nabla}\psi^{(\lambda,u)}&=i\lambda \psi^{(\lambda,u)},
\end{align}
$\mu^{(\frac{1}{2})}_b(\lambda)$ is the Plancherel measure of the spin-half field on $AdS_b$ \cite{Camporesi:1995fb}
\begin{align}
\mu^{(\frac{1}{2})}_b(\lambda)= \frac{\pi}{2^{2b-4}\Gamma(\frac{b}{2})^2}
\begin{cases}
    \prod_{i=1}^{\frac{b-2}{2}}(\lambda^2+i^2) & \qquad \rm{ Odd}\hspace{2mm} b,\\
     \lambda\coth\pi\lambda \prod_{i=1}^{\frac{b-2}{2}}(\lambda^2+i^2)&\qquad \rm{ Even} \hspace{2mm} b.
\end{cases}
\end{align}
We perform the integral over $\lambda$ by using Hubbard-Stratonovich  trick and obtain
\begin{align}\label{spinhalfstep1}
    -\mathcal{F}^{\frac{1}{2}}_{S^a\times AdS_b}&=\frac{1}{4}\int_{C}du\int_0^{\infty}\frac{d\tau}{\sqrt{4\pi\tau^3}} \sum_{n=0}^{\infty}g^{(\frac{1}{2})}_n e^{-\frac{\epsilon^2+u^2}{4\tau}}e^{-\tau(n+\frac{a}{2})^2}W_{\frac{1}{2}}^{(b)}(u).
\end{align}
$W_{\frac{1}{2}}^{(b)}(u)$ is the Fourier transform of the Plancherel  measure of the spin-half field on $AdS_b$ \cite{Sun:2020ame}.
\begin{align}
W_{\frac{1}{2}}^{(b)}(u)&=(-1)^b2^{b}\frac{e^{-\frac{b}{2}u}}{(1-e^{-u})^b}. 
\end{align}
Here the contour  $\rm{C_{odd}}$ or $\rm{C_{even}}$ is given in figure [\ref{fig4}] and figure [\ref{fig5}]
 depending on whether $b$ is odd or even respectively. 
We perform the  integral over $\tau$ and obtain
\begin{align}\label{KKsumspin half}
   -    \mathcal{F}^{\frac{1}{2}}_{S^a\times AdS_b}&=\int_C\frac{du}{4\sqrt{u^2+\epsilon ^2}}\sum_{n=0}^{\infty}g^{(\frac{1}{2})}_ne^{  \left(-(n+\frac{a}{2})\sqrt{u^2+\epsilon ^2}\right)}W_{\frac{1}{2}}^{(b)}(u).
\end{align}

 For the case when $b$ is even substituting $W_{\frac{1}{2}}^{(b)}$  and using the 
 contour $\rm{C_{even}}$ as shown in figure [\ref{fig4}] .
 We perform the sum over eigen modes on $S^a$ to  obtain
\begin{align}
  - \mathcal{F}^{\frac{1}{2}}_{S^a\times AdS_b}& = -\int_{R+i\delta}\frac{du}{4\sqrt{u^2+\epsilon ^2}}\frac{e^{-\frac{u}{2}}}{(1-e^{-u})}\frac{2^{a-1}e^{-u\frac{a-1}{2}}}{(1-e^{-u})^{a-1}}(-1)^b2^{b}\frac{e^{-\frac{b}{2}u}}{(1-e^{-u})^b}\nonumber\\
   &=-\int_{R+i\delta}\frac{du}{2u}\frac{2e^{-\frac{u}{2}}}{(1-e^{-u})}\frac{2^{d-1}e^{-u\frac{d-1}{2}}}{(1-e^{-u})^{d-1}}
\end{align}
In the last line we have substituted $a+b=d$.\\
Similarly we proceed for odd $b$ where the contour is given in \ref{fig5}. 
\begin{align}
    \mathcal{F}^{\frac{1}{2}}_{S^a\times AdS_b}&=-\frac{\log R}{2\pi i}\int_{C_{\rm{Odd}}}\frac{du}{2u}\frac{2e^{-\frac{u}{2}}}{(1-e^{-u})}\frac{2^{d-1}e^{-u\frac{d-1}{2}}}{(1-e^{-u})^{d-1}}.
\end{align}
Therefore we obtain the partition function of the massless Dirac fermion on $S^a\times AdS_b$ which has the same expression as we obtained it for $S^{a+b}$ in \eqref{spinhalfspherepartition}. In the recent work of \cite{Sato:2021eqo} the equality of the partition function of Dirac fermion has been checked numerically. From the experience of the scalar field, it is expected that the partition function of the free fermion should also be the same in the classes of conformally equivalent spaces. Since we express the partition function as an integral transform of the Harish-Chandra character the equality of the partition functions becomes explicit. Therefore the anomaly coefficients are also the same in the classes of conformally equivalent spaces.
\subsection{\texorpdfstring{$\slashed{\partial}^3$}{TEXT} operator}
The action of Weyl invariant $3$-derivative operator for the Majorana fermion can be found \cite{Gilkey:1975iq} and in $d=4$ it is given by \cite{Beccaria:2017dmw,Fradkin:1981jc}
\begin{align}
    S=i\int d^4x \hspace{1mm}\sqrt{g}\left(\Bar{\psi}\big[\slashed{\nabla}^3+(R^{\mu\nu}-\frac{1}{6}g^{\mu\nu}R)\gamma_{\mu}\nabla_{\nu}\big]\psi\right),
\end{align}
where $R^{\mu\nu}$ is the Ricci tensor in four dimension.
\subsection*{$\slashed{\partial}^3$ operator on $S^4$}We evaluate the partition function of the $3$-derivative fermionic fields on $S^4$.
The Ricci tensor and the curvature are known
\begin{align*}
    R^{\mu\nu}=\frac{R}{4}g^{\mu\nu},\qquad R=12.
\end{align*}
We now substitute the curvature use the method of factorisation in Weyl invariant $3$-derivative fermionic operator to obtain
\begin{align}\label{ferm3der}
    \mathcal{O}^{(3)}_{\frac{1}{2}}&=\slashed{\nabla}^3+(R^{\mu\nu}-\frac{1}{6}g^{\mu\nu}R)\gamma_{\mu}\nabla_{\nu}\nonumber\\
    &=\slashed{\nabla}(\slashed{\nabla}^2+1).
\end{align}
Therefore the free energy can be written as
\begin{align}
    \mathcal{F}^{(3)}_{\frac{1}{2}}&=-\rm{Tr}\log(i \mathcal{O}^{(3)}_{\frac{1}{2}}).
\end{align}
One evaluates the free energy of the fermionic fields by using the spectrum of the squared of the operator.\\
We use the identity \eqref{logiden} and substitute the spectrum of the squared operator $ \mathcal{O}^{(3)}_{\frac{1}{2}}$ to obtain
\begin{align}
  -\mathcal{F}^{(3)}_{\frac{1}{2}}&  =\int_{\epsilon}^{\infty}\frac{d\tau}{2\tau} e^{-\frac{\epsilon^2}{4\tau}}\sum_{n=0}^{\infty}\left(g_{n,4}^{(\frac{1}{2})} e^{i u(n+2)}(1+2e^{-\tau})\right)
\end{align}
Note that we have dropped the second term of the logarithm using the similar argument presented in \eqref{spinhalfsum}. Therefore we start with a convergent integral where the prescription of `dimensional regularisation' has been used to sum over degeneracy \cite{Klebanov:2011gs}. But we still introduce the factor $e^{-\frac{\epsilon^2}{4\tau}}$ to keep track of the branch cut in the integration plane \cite{David:2021wrw} and take $\epsilon\rightarrow 0$ in the end. \\
Now we follow the same steps from \eqref{spinhalfstepinitial} to \eqref{spinhalfspherepartition} and obtain
\begin{align}\label{3derspinorsphere}
 -\mathcal{F}^{(3)}_{\frac{1}{2}}& =\int_0^{\infty}\frac{dt}{2t}\frac{2e^{-\frac{t}{2}}}{1-e^{-t}}    4\left(\frac{e^{-\frac{t}{2}}+e^{-3\frac{t}{2}}+e^{-5\frac{t}{2}}}{(1-e^{-t})^3}\right).
\end{align}
Therefore we express the partition function of $3$-derivative spin-half field in terms of character integral and the log divergent part can be extracted by expanding the integrand \eqref{3derspinorsphere} around $t=0$ and collecting the coefficient of $1/t$ term.
\subsection*{$\slashed{\partial}^3$ operator on $S^1_q\times AdS_3$}
Branched sphere $S^4_q$ is conformally related to the hyperbolic cylinder $AdS_3\times S^1_q$. We, therefore, evaluate the partition function of $3$-derivative Weyl invariant fermionic field on the hyperbolic cylinder and compare it with the sphere at $q=1$.\footnote{ As far as we know the spectrum and the degeneracy of the spin-half field are not known on the branched sphere.} 
In $S^1_q\times AdS_3$ the $3$-derivative Weyl invariant operator has a different form because the curvature induced mass comes only from the $AdS_3$ space,
\begin{align*}
    R_{00}=0, \qquad R_{oi}=0, \qquad R_{ij}=-2g_{ij}.
\end{align*}
Here $i,j$ denotes the direction along $AdS_3$. Therefore the operator $\mathcal{O}^{(3)}_{\frac{1}{2}}$ becomes \cite{Beccaria:2017dmw}
\begin{align}
    \mathcal{O}^{(3)}_{\frac{1}{2}}=\slashed{\nabla}^3+\gamma^0\partial_0-\gamma^i\nabla_i.
\end{align}
To obtain free energy one uses the spectrum of the squared operator. Therefore we square the operator $\mathcal{O}^{(3)}_{\frac{1}{2}}$ and factorize it 
\begin{align}
    (i\mathcal{O}^{(3)}_{\frac{1}{2}})^2=(-\partial_0^2-\slashed{\nabla}^2)(-\partial_0^2+(i\slashed{\nabla}+i)^2)(-\partial_0^2+(i\slashed{\nabla}-i)^2).
\end{align}
Free energy becomes
   \begin{align}
   -\mathcal{F}^{(3)}_{\frac{1}{2}}&= \frac{1}{4}\sum_{n\in \mathbb{Z}+\frac{1}{2}}^{\infty}\int_{-\infty}^{\infty}\mu^{(\frac{1}{2})}_3(\lambda)\bigg[\log\left(\lambda^2+\frac{n^2}{q^2}\right)+\log\left((\lambda+i)^2+\frac{n^2}{q^2}\right)+\log\left((\lambda-i)^2+\frac{n^2}{q^2}\right)\bigg].
\end{align}
Note that we sum over half integers due to the anti-periodicity of the fermions along $S^1_q$ direction. $\mu^{(\frac{1}{2})}_3(\lambda)$ is the Plancherel measure of spin-half field on $AdS_3$.\\
We now follow the same steps from \eqref{spinhalfstep1} to \eqref{KKsumspin half} where we use the Hubbard -Stratonovich trick and sum over the Kaluza-Klein modes to obtain
\begin{align}
    -\mathcal{F}^{(3)}_{q,\frac{1}{2}}&=\frac{\log R}{2\pi i} \int_{\rm{C_{Odd}}}\frac{du}{2u}\frac{8 e^{-3\frac{u}{2}}\sum_{n\in\mathbb{Z}+\frac{1}{2}}e^{-u\frac{n}{q}}}{(1-e^{-u})^3}(1+e^{-u}+e^{u})\nonumber\\
    &=\frac{\log R}{2\pi i} \int_{\rm{C_{Odd}}}\frac{du}{2u}\frac{8 \left(e^u+e^{2 u}+1\right) e^{\frac{(q+1) u}{2 q}}}{\left(e^u-1\right)^3 \left(e^{u/q}-1\right)}.
\end{align}
At $q=1$ we obtain
\begin{align}\label{derspinorhyp}
     -\mathcal{F}^{(3)}_{1,\frac{1}{2}}=\frac{\log R}{2\pi i} \int_{\rm{C_{Odd}}}\frac{du}{2u}\frac{2e^{-\frac{u}{2}}}{1-e^{-u}}    4\left(\frac{e^{-\frac{u}{2}}+e^{-3\frac{u}{2}}+e^{-5\frac{u}{2}}}{(1-e^{-u})^3}\right).
\end{align}
The log divergent part of the free energy on hyperbolic cylinder can be evaluated by taking the residue of the integral \eqref{derspinorhyp} at $u=0$. Note that the log divergent piece of the free energies are identical given in \eqref{3derspinorsphere} and \eqref{derspinorhyp}. Therefore the anomaly coefficient of $3$-derivative spin-half field also agrees.
We evaluate the log divergent part of the free energy and the universal part of the R\'{e}nyi entropy can be extracted from it.
\begin{align}
    \mathcal{F}^{(3)}_{q,\frac{1}{2}}=-\frac{-29 q^4-50 q^2+7}{480 q^3}, \qquad S_q=-\frac{(q+1) \left(43 q^2-7\right)}{480 q^3}.
\end{align}
The R\'{e}nyi entropy agrees with \cite{Beccaria:2017dmw}.
We also evaluate the conformal dimension of the twist operator from the free energy of $3$-derivative fermionic field on hyperbolic cylinder.
\begin{align}\label{hqfer}
        h_q&=\frac{q}{3 \rm{Vol}(AdS_3)}\left(\partial_q\mathcal{F}^{(3)}_{q,\frac{1}{2}}|_{q=1}-\partial_q\mathcal{F}^{(3)}_{q,\frac{1}{2}}\right)\nonumber\\
        &=-\frac{29 q^4-50 q^2+21}{2880 \pi  q^3}.
    \end{align}
    The first and second derivative of $h_q$ with respect to $q$ at $q=1$ can be related to the two and three point functions of stress tensor which will be useful to understand the causality bound of the $3$-derivative fermionic fields.
\section{ Vector field}
In this section, we examine the Weyl-covariant vector field. The partition functions of Weyl-invariant scalar and fermions are identical on conformally related spaces. In \cite{David:2021wrw} it was demonstrated that the hyperbolic cylinder misses out the edge mode of the partition function of conformal $p$-form fields.  We evaluate the partition function of the higher derivative Weyl-covariant vector field on the branched sphere as well as on the hyperbolic cylinder to find out whether the same pattern is followed.
\begin{subsection}{Maxwell field in \texorpdfstring{$d=4$}{TEXT}}
It is well known that the free Maxwell field is conformal in $d=4$ dimension which has been studied extensively and the universal part of the entanglement entropy across a spherical entangling surface has been evaluated \cite{Soni:2016ogt}. Since the theory is conformal in four dimensions one can use conformal mapping to study the partition function and entanglement entropy of it across a spherical surface.
The partition function of free Maxwell field on $S^4_q$ differs from the $AdS_3\times S^1_q$ by edge mode which is a partition function of massless scalar field on $S^2$ \cite{David:2021wrw, Huang:2014pfa,Donnelly:2014fua}. In \cite{Benedetti:2019uej} it was shown that the universal part of the entanglement entropy can be evaluated from the canonical quantization of the field on a sphere which results in $2$ independent scalar modes with angular momentum $\ell\geq 1$ and it agrees with the hyperbolic cylinder computation. Therefore one obtains the `extractable' part of the entanglement entropy \cite{Soni:2016ogt} of free Maxwell field  from the partition function of $AdS_3\times S^1_q$.
\subsection*{Maxwell field on $S^4_q$} In this section we review the method to cast the partition function of vector field in terms of the Harish-Chandra character on $S^4_q$. The gauge fixed partition function of the $p$-form field on  sphere is given in \cite{Obukhov:1982dt,Copeland:1984qk,Cappelli:2000fe}. The partition function of vector field on branched sphere in $4$-dimension is given by \cite{Beccaria:2017lcz}
\begin{align}
    \mathcal{Z}[S^4_q]^{(1)}&=\left(\frac{\det(-\Delta_0)}{\det(-\Delta_1+3)}\right)^{\frac{1}{2}}
\end{align}
Here $\Delta_0$ and $\Delta_1$ are the spin-$0$ and spin-$1$ Laplacian  on $S^4_q$ respectively.\\
We start with the transverse spin-$1$ part of the partition function.
 The eigen-value and corresponding degeneracy are known \cite{DeNardo:1996kp}
 \begin{align}
     \begin{split}
     \lambda^{(1)}_{n,m}&=(n+\frac{m}{q})(n+\frac{m}{q}+3)-1,  \qquad \hbox{with} \; n+m \geq 1,  \; n, m \in\{ 0, 1, 2, \cdots\}\nonumber\\
     g^{(1)}_{n,m=0}&=\frac{1}{n+1}   \binom{n+1}{n-1}
   \{3n+5  \},  \qquad n = 1, 2, \cdots , 
   \\ \nonumber
        g^{(1)}_{n,m}&=6 \binom{n+3}{2}, \qquad n = 0, 1, \cdots \qquad m =  1, 2, \cdots
     \end{split}
 \end{align}
 The logarithm of determinant of transverse spin-$1$ field with curvature induced mass is therefore given by 
\begin{eqnarray}
-\frac{1}{2}\log {\rm det}(-\Delta_1+3 )   = -\frac{1}{2} 
\sum_{n,m =1}^\infty g_{n,m }^{(1)}  \log ( \lambda_{n,m}^{(1)}+3 ) .
\end{eqnarray}
 Using the identity (\ref{logiden})  we rewrite the partition function as
 \begin{align}\label{vectorstart}
     -\frac{1}{2}\log {\rm det}(-\Delta_1+3 ) =\int_0^\infty \frac{d\tau}{2\tau} 
\sum_{n,m =1}^\infty g_{n,m}^{(1)} ( e^{-\tau (\lambda_{n, m}^{(1)}+3) } - e^{-\tau} ) .
 \end{align}
 Now the second term involves the sum over degeneracies. We regulate it by looking at the sufficiently large negative $d$ and use the zeta regularization prescription. 
We obtain 
\begin{equation}\label{zetareg}
\sum_{n =1}^\infty \sum  g_{n, m=0}^{(1)}  = 1, 
\qquad\qquad
\sum_{n =0}^\infty\sum_{m =1}^\infty g_{n, m}^{(1)} =0.
\end{equation}
Note that \eqref{vectorstart} is already convergent since the second term has been regularised. Therefore $\epsilon$ helps us to keep track of the branch cut in the integration plane. One can notice that
\begin{eqnarray} 
g_{0, m =0}^{(1)} = 0,  \qquad \qquad g_{-1, m =0}^{(1)}  = 1, \qquad \lambda_{-1, m =0}^{(1)} = 0 .
\end{eqnarray}
Therefore we rewrite the partition function
\begin{align}
-\frac{1}{2}\log {\rm det}(-\Delta_1+3 )   =  \int_0^\infty \frac{d\tau}{2\tau} 
e^{-\frac{\epsilon^2}{ 4\tau} }  \left( \sum_{n =-1}^\infty g_{n, m =0}^{(1)}  e^{-(\lambda_{n, m =0}^{(1)} +3)\tau}
+ \sum_{n =0, m >0}^\infty g_{n, m }^{(1)} e^{-(\lambda_{n, m}^{(1)}+3) \tau } \right),
\end{align}
We follow the same steps discussed for scalar field from \eqref{Hubbard} to \eqref{u part} and take $\epsilon\rightarrow 0$ in the end
\begin{eqnarray}
    -\frac{1}{2}\log {\rm det}(-\Delta_1+3 )& =&\int_0^{\infty}\frac{dt}{2t} \left\{ 
    \frac{1+e^{-\frac{t}{q} } }{1-e^{-\frac{t}{q}} } \left[ 
   \binom{3}{1}
   \frac{e^{-t}+e^{-2t} }{(1-e^{-t})^3} \right]  \right.  \\ \nonumber
   & &  \qquad\qquad\qquad 
    \left.   - \frac{1+e^{-t} }{1-e^{-t} }  \left[   \binom{1}{0}\frac{1+e^{-t} }{ ( 1- e^{ -t} ) } 
   \right]  \right\}.
\end{eqnarray}
Similarly we evaluate the massless scalar part of the partition function and obtain
\begin{align}
    -\frac{1}{2}\log {\rm det}(-\Delta_0 )&=\int_0^{\infty}\frac{dt}{2t} \frac{1+e^{-\frac{t}{q}}}{1-e^{-\frac{t}{q}}}\binom{3}{0}\frac{1+e^{-3t}}{(1-e^{-t})^3}.
\end{align}
Therefore combining the transverse spin-1 part and the determinant of the massless scalar in the numerator we get 
\begin{align}\label{partvectorsphere}
\log \mathcal{Z}[S^4_q]^{(1)}&=\int_0^{\infty}\frac{dt}{2t}\left(\frac{1+e^{-\frac{t}{q}}}{1-e^{-\frac{t}{q}}}\hat{\chi}^b_1-\frac{1+e^{-t}}{1-e^{-t}}\hat{\chi}^e_1\right)
\end{align}
Here $\hat{\chi}^b_1$ is the full `naive' bulk character  and $\hat{\chi}^e_1$ is the corresponding `naive' edge character of the free Maxwell field on $S^4$. 
\begin{align}
    \begin{split}
\hat{\chi}^b_1&=     \binom{3}{1}
   \frac{e^{-t}+e^{-2t} }{(1-e^{-t})^3}   -\binom{3}{0}\frac{1+e^{-3t}}{(1-e^{-t})^3},\\
   \hat{\chi}^e_1&=\binom{1}{0}\frac{1+e^{-t} }{ ( 1- e^{ -t} )}.
    \end{split}
\end{align}
Therefore we have obtained the `naive' characters of the vector field in $d=4$ dimensions in the sense of \cite{Anninos:2020hfj}. Now to obtain the `flipped' character we subtract the term which contributes at $t\rightarrow\infty$. Therefore the full bulk character becomes
\begin{align}
    \chi^b= \binom{3}{1}
   \frac{e^{-t}+e^{-2t} }{(1-e^{-t})^3}   -\binom{3}{0}\frac{1+e^{-3t}}{(1-e^{-t})^3}+1.
\end{align}
Note that the `flipping' of the character does not alter the co-efficient of $1/t$ for an even dimension. So we can still use the small $t$ expansion about $t=0$ and collect the log divergent part of the partition function.
The universal part of the free energy therefore becomes \cite{David:2021wrw}
\begin{align}
    \mathcal{F}[S^4_q]^{(1)}=\frac{33 q^4+30 q^2+1}{180 q^3}+\frac{1}{3}., \qquad \mathcal{F}[S^4_q]^{(1)}|_{q=1}=\frac{31}{45}.
\end{align}
The log divergent part of free energy of masses spin-$1$ field $\frac{31}{45}$ agrees with \cite{Beccaria:2017lcz, Anninos:2020hfj} which is also known as the anomaly coefficient. Therefore one obtains the anomaly coefficient of the Maxwell field from the partition function on $S^4$.
\subsection*{Maxwell field on $AdS_3\times S^1_q$}
We have obtained the anomaly coefficient of the Maxwell field from the partition function on $S^4$. It is well known that $S^4_q$ and $AdS_3\times S^1_q$ are related by Weyl transformation. Therefore it will be interesting to evaluate the log divergent part of the free energy of Maxwell field on $AdS_3\times S^1_q$ and compare it with the $S^4_q$.\\
In this section we review the partition function of the free Maxwell field on $AdS_3\times S^1_q$ \cite{David:2020mls,Beccaria:2017dmw}. In \cite{David:2020mls} gauge fixed partition functions of $1$-form and $2$-form on hyperbolic cylinder were obtained and the method was extended to  arbitrary $p$-form. The gauge fixing method was adapted from \cite{Fradkin:1983mq}.\\
The partition function of the vector field on $AdS_3\times S^1_q$ is given by
\begin{align}
     \mathcal{Z}[AdS_3\times S^1_q]^{(1)}&=\left[\frac{1}{\det(-\partial_{\tau}^2-\Delta_{p=1})}\right]^{\frac{1}{2}}.
\end{align}
Here $\Delta_{p=1}$ is the Hodge-deRham Laplacian acting on co-exact 1-form field on $AdS_3$.  The operator $\partial_\tau^2$ 
picks out the Kaluza-Klein mass along the $S^1$ direction.
The eigen value of the  Hodge-deRham Laplacian on 1-form on $AdS_3$ is given by \cite{Camporesi:1994ga}
\begin{equation} \label{eigenhdr}
\Delta_{p=1}\psi^{\{ \lambda, u\}}_{i}= 
- \lambda^2   
 \psi^{\{ \lambda, u\}}_{i}.
\end{equation}
Here $\psi^{\{ \lambda, u\}}_{i}$ refers  to the eigen functions and $\{u\}$ denote the other quantum numbers on $AdS_3$. 
Using the mode expansion along $S^1_q$ and substituting the eigen-values we obtain
\begin{equation}
\log  \mathcal{Z}[S^1_q\times AdS_3]^{(1)} = 
-\frac{1}{2} \sum_{n=-\infty}^\infty \int d \mu_1^{(3)} (\lambda)\log \left( 
\frac{n^2}{q^2} + \lambda^2 \right) .
\end{equation}
where $\mu_1^{(3)} (\lambda)$ is the Plancherel measure of the co-exact $1$-form field on $AdS_3$ \cite{Camporesi:1994ga}
\begin{align}
 d\mu_1^{(3)} (\lambda)=\frac{\rm{Vol}(AdS_3)}{\pi^2}(\lambda^2+1)   .
\end{align}
Now we substitute this back in the expression of the partition function and follow the same steps shown for the scalar on hyperbolic cylinder we obtain \cite{David:2020mls}
\begin{align} \label{parthypvector}
    \log \mathcal{Z}[S^1_q\times AdS_3]^{(1)} &=\frac{1}{ 2\pi i } \log R
\int_{C_{\rm{Odd}}} \frac{dt}{2t}\frac{1+e^{-\frac{t}{q}}}{1-e^{-\frac{t}{q}}}\left( \binom{3}{1}
   \frac{e^{-t}+e^{-2t} }{(1-e^{-t})^3}   -\binom{3}{0}\frac{1+e^{-3t}}{(1-e^{-t})^3}\right).
\end{align}
We now compare \eqref{parthypvector} and \eqref{partvectorsphere} and observe that the partition function of conformal vector field on $AdS_3\times S^1_q$ misses out the edge mode part of the partition function. Note that the `edge' mode part of the partition function does not depend on the branching parameter $q$. We obtain the log divergent part of the free energy by taking residue of the integral \eqref{parthypvector} at $t=0$.
\begin{align}
    \mathcal{F}_q[S^1_q\times AdS_3] ^{(1)}=\frac{33 q^4+30 q^2+1}{180 q^3}, \qquad  \mathcal{F}_{q=1}[S^1_q\times AdS_3] ^{(1)}=\frac{16}{45}.
\end{align}
The free energy on hyperbolic cylinder does not capture the 'edge' mode contribution of $\frac{1}{3}$ which comes partition function of massless scalar on $S^2$.
This behavior was noticed earlier in the study of the universal part of the entanglement entropies \cite{Casini:2013rba, Donnelly:2015hxa, Donnelly:2014fua, Huang:2014pfa, Casini:2015dsg} of the Maxwell field in $d=4$ dimensions, and the `extractable' part of the entanglement entropy comes from the free energy of the hyperbolic cylinder.  In \cite{David:2021wrw} it was explicitly shown that the partition function of  conformal $p$-forms on hyperbolic cylinder contains only the `bulk' mode and misses out the 'edge' mode.  Therefore one expects the `extractable' part of the entanglement entropy across a spherical region can be obtained from the partition function of the hyperbolic cylinder which misses out the `edge' mode.
\end{subsection}
\subsection{Conformal higher derivative vector field}In the previous section we have discussed the techniques to evaluate the partition function of the free Maxwell field in $d=4$ dimension and demonstrated that the partition function of the hyperbolic cylinder differs from the branched sphere by edge mode which is a partition function of $0$-form field on $S^2$. To understand this feature we evaluate the partition function of the Weyl-covariant higher derivative vector gauge field on the branched sphere as well on the hyperbolic cylinder. We compare the log divergent part of the free energy of higher derivative conformal vector gauge field on $S^6_q$ and $AdS_5\times S^1_q$ to understand the edge mode characteristics in detail.
The action of Weyl-invariant vector gauge theory in  6-dimension  is given by \cite{Beccaria:2017dmw}
\begin{align}\label{actionvect}
    S=\int \sqrt{g}d^6x \left(\nabla_{\lambda}F^{\lambda\mu}\nabla^{\nu}F_{\nu\mu}-(R_{\mu\nu}-\frac{R}{5}g_{\mu\nu})F^{\mu\lambda}F^{\nu}_{~\lambda}\right),
\end{align}
where $F_{\mu\nu}=\partial_{\mu}A_{\nu}-\partial_{\nu}A_{\mu}$ and the action is invariant under gauge transformation $A_{\mu}\rightarrow A_{\mu}+\nabla_{\mu}\epsilon$. 
\subsection*{ Four-derivative Weyl-covariant vector field on $S^6_q$}We evaluate the partition function of the Weyl-covariant vector gauge theory on a branched sphere in 6 dimension. We follow the gauge fixing method which was introduced in \cite{Fradkin:1983mq}.
We start with decomposing the field variable into transverse mode and longitudinal mode in the following way
\begin{eqnarray}
A_{\mu}=\{A^{\perp}_{\mu}+\nabla_{\mu}\phi\},\qquad\rm{where}\qquad \nabla_{\mu}A^{\perp\mu}=0.
\end{eqnarray}
Under the gauge transformation transverse mode does not change but the longitudinal mode gets shifted by constant.
\begin{eqnarray}
A^{\perp\mu}\rightarrow A^{\perp\mu},\qquad\phi\rightarrow\phi+\epsilon
\end{eqnarray}
Substituting the decomposed variables in the action
\begin{align}
    S&=\int \sqrt{g}d^6x\big[\nabla^2A^{\perp}_{\mu}-R_{\rho\mu}A^{\perp\rho})(\nabla^{2}A^{\perp\mu}-R_{\rho\mu}A^{\perp\rho})\nonumber\\
    &-2(\frac{R}{6}-\frac{R}{5})(-A^{\perp}_{\mu}\nabla^2A^{\perp\mu}
    +\frac{R}{6}A^{\perp\mu}A^{\perp}_{\mu}\big]\nonumber\\
    &=\int \sqrt{g}d^6x  A^{\perp}_{\mu}\left(-\Delta_{(1)}+5\right)\left(-\Delta_{(1)}+7\right)A^{\perp\mu},
\end{align}
where $\nabla^2=\nabla_{\mu}\nabla^{\mu}=\Delta_{(1)}$ is the spin-1 Laplacian on $S^6_q$.
We have used the integration by parts and dropped terms using transverse gauge condition $\nabla_{\mu}A^{\perp\mu}=0$. Covariant derivative operators are interchanged using
\begin{eqnarray}
[\nabla_{\mu},\nabla_{\nu}]A^{\perp\lambda}=R^{\lambda}_{\rho\mu\nu}A^{\perp\rho}, \qquad R_{\mu\nu\rho\sigma}=(g_{\mu\rho}g_{\nu\sigma}-g_{\mu\sigma}g_{\nu\rho}).
\end{eqnarray}
The second line comes from the fact that $S^6_q$ is a maximally symmetric space with unit radius. Since we have changed the field variables we will get a change in measure
\begin{align}
    \mathcal{D}A^{\mu}=(\det(-\Delta_0)^{\frac{1}{2}}) \mathcal{D}A^{\perp\mu}\mathcal{D}\phi,
\end{align}
where $(-\Delta_0)$ is the spin-0 Laplacian on $S^6_q$. Note that the action is independent of the longitudinal mode and hence it is gauge invariant under
$$\phi\rightarrow \phi+\epsilon.$$  Therefore we can fix the gauge by choosing the gauge slice 
$$\mathcal{G}=\delta(\phi).$$
But this gauge slice corresponds to the unit Fadeev-Popov determinant. Therefore the partition function can be written as \cite{Beccaria:2015uta}
\begin{align}
    \mathcal{Z}^{(4)}_{\rm{v}}&=\left(\frac{\det(-\Delta_0)}{\det\left(-\Delta_{(1)}+5\right)\det\left(-\Delta_{(1)}+7\right)}\right)^{\frac{1}{2}}.
\end{align}
In the last step we have integrated the transverse components of the gauge field and used the Fadeev-Popov determinant to write the complete partition function.\\
The eigen-value and the degeneracy of the spin-1 field on $S_q^6$ is given by \cite{DeNardo:1996kp}
\begin{equation}
   \lambda^{(1)}_{n,m}=(n+\frac{m}{q})(n+\frac{m}{q}+5)-1 ,  \qquad \hbox{with} \; n+m \geq 1,  \; n, m \in\{ 0, 1, 2, \cdots\}
\end{equation}
\begin{eqnarray}\label{bsdegen}
   g^{(1)}_{n,m=0}&=&\frac{1}{24} n (n+2) (n+3) (5 n+17),  \qquad n = 1, 2, \cdots , 
   \\ \nonumber
        g^{(1)}_{n,m}&=&\frac{5}{12} (n+1) (n+2) (n+3) (n+4), \qquad n = 0, 1, \cdots \qquad m =  1, 2, \cdots 
\end{eqnarray}
Note that eigen-value and degeneracy on branched sphere are labelled by $2$ integers.\\
Now we evaluate
\begin{eqnarray}\label{step1}
-\frac{1}{2}\log \det\left(-\Delta_{(1)}+5\right)   = -\frac{1}{2} 
\sum_{n, m =0}^\infty g_{n, m }^{(1)}  \log ( \lambda_{n, m}^{(1)}+5 ) .
\end{eqnarray}
Again using the identity (\ref{logiden})  we rewrite \eqref{step1} as
\begin{eqnarray} \label{bs1f}
-\frac{1}{2}\log\det\left(-\Delta_{(1)}+5\right)  =  \int_0^\infty \frac{d\tau}{2\tau} 
\sum_{n, m =0}^\infty g_{n, m}^{(1)} ( e^{-\tau (\lambda_{n, m}^{(1)}+5) } - e^{-\tau} ) .
\end{eqnarray}
The second term involves the sum over degeneracy which we drop by the same logic presented in \eqref{zetareg}.
Therefore the equation (\ref{bs1f}) reduces to 
\begin{equation} \label{bs1f12}
-\frac{1}{2}\log\det\left(-\Delta_{(1)}+5\right) =    \int_0^\infty \frac{d\tau}{2\tau} 
e^{-\frac{\epsilon^2}{ 4\tau} } \left(  \sum_{n , m =0}^\infty g_{n, m }^{(1)}e^{-\tau(\lambda_{n, m}^{(1)}+5)}  - 1 \right) ,
\end{equation}
 The second term in (\ref{bs1f12})  can be absorbed in the first term by 
noting from (\ref{bsdegen}) that 
\begin{eqnarray} 
g_{0, m =0}^{(1)} = 0,  \qquad \qquad g_{-1, m =0}^{(1)}  = 1, \qquad \lambda_{-1, m =0}^{(1)} = 0 .
\end{eqnarray}
Here $\epsilon$ is used to keep track of track of the branch cut in the integration plane.\\
Following the steps given from \eqref{contint} to \eqref{endstep} we obtain
\begin{align} \label{bs1f1}
-\frac{1}{2}\log\det\left(-\Delta_{(1)}+5\right)& =    \int_{\epsilon}^{\infty}\frac{dt}{2 \sqrt{t^2-\epsilon^2}}(e^{\frac{3}{2} \sqrt{t^2-\epsilon^2}}+
e^{-\frac{3}{2}\sqrt{t^2-\epsilon^2}})(f_1(it)+f_2(it)).
\end{align}
Similarly we evaluate 
\begin{align}
    -\frac{1}{2}\log\det\left(-\Delta_{(1)}+7\right)& =    \int_{\epsilon}^{\infty}\frac{dt}{2 \sqrt{t^2-\epsilon^2}}(e^{\frac{1}{2} \sqrt{t^2-\epsilon^2}}+
e^{-\frac{1}{2}\sqrt{t^2-\epsilon^2}})(f_1(it)+f_2(it)).
\end{align}
and
\begin{align}
     -\frac{1}{2}\log\det\left(-\Delta_0\right)& =    \int_{\epsilon}^{\infty}\frac{dt}{2 \sqrt{t^2-\epsilon^2}}(e^{\frac{5}{2} \sqrt{t^2-\epsilon^2}}+
e^{-\frac{5}{2}\sqrt{t^2-\epsilon^2}})(f_0(it)+f_0(it)).
\end{align}
where\begin{eqnarray}
  f_1(u)&=&\sum_{n=-1}^{\infty}g^{(1)}_{n,m=0}e^{iu((n+\frac{5}{2})}
        =\frac{e^{-\frac{1}{2} (3 i u)} \left(5 e^{-i u}-e^{-2 i u}+1\right)}{\left(-1+e^{-i u}\right)^5},\nonumber\\
        f_2(u)&=&\sum_{m=1, n=0}^{\infty} g^{(1)}_{n,m}e^{iu((n+\frac{m}{q} +\frac{5}{2})} 
        =\frac{10 e^{-\frac{1}{2} (5 i u)}}{\left(-1+e^{-i u}\right)^5 \left(-1+e^{-\frac{i u}{q}}\right)}.\nonumber\\
       f_0 ( u ) &=& \sum_{n,m =0}^\infty g^{(0)}_{n,,m}e^{i(n+\frac{m}{q}+\frac{5}{2})u}
    = \frac{e^{ \frac{i u 5}{2}}}{ ( 1 - e^{i u } )^5 } \frac{ 1 + e^{ i \frac{ u }{ q}} }{ 1 - e^{ i \frac{ u }{ q}}}.
\end{eqnarray}
Now we substitute these functions  in the partition function and take $\epsilon\rightarrow 0$ in the end to obtain
\begin{eqnarray}\label{spherevector}
\log\mathcal{Z}^{(4)}_v&=&\int_0^{\infty}\frac{dt}{2t}\left\{ 
    \frac{1+e^{-\frac{t}{q} } }{1-e^{-\frac{t}{q}} } \left[ 
   (5\frac{(e^{-t}+e^{-4t})}{(1-e^{-t})^5}+5\frac{(e^{-2t}+e^{-3t})}{(1-e^{-t})^5}-\frac{(1+e^{-5t})}{(1-e^{-t})^5} \right]  \right. \nonumber\\
   & &  \qquad\qquad\qquad 
     \left.   - \frac{1+e^{-t} }{1-e^{-t} }  \left[   \frac{e^{-t}+e^{-2t} }{ ( 1- e^{ -t} )^3 } +\frac{1+e^{-3t}}{(1-e^{-t})^3}
   \right]  \right\} \nonumber\\
   &=&\int_0^{\infty}\frac{dt}{2t}\left\{ 
    \frac{1+e^{-\frac{t}{q} } }{1-e^{-\frac{t}{q}}}\hat{\chi}_{\rm{bulk}}- \frac{1+e^{-t} }{1-e^{-t} }\hat{\chi}_{\rm{edge}}\right\}.
\end{eqnarray}
 $\hat{\chi}_{\rm{bulk}}$ and $\hat{\chi}_{\rm{edge}}$ correspond to the full `naive' bulk character and `naive' edge character of four-derivative conformal vector field on $S^6_q$ respectively. Note that the `naive' edge character comes with a kinematic factor which is independent of the branching parameter $q$ which is similar to the vector case on $S^4_q$. Here we wish to mention that these `naive' character do not correspond to the $\rm{UIR}$ of exceptional series. It will be interesting to check that the `flipping' procedure introduced in \cite{Anninos:2020hfj} leads to the $\rm{UIR} $ of any exceptional series. But here we content ourselves with the `naive' character because we are interested in evaluating the universal piece of the entanglement entropy which comes from the logarithmic divergent part of the partition function and it does not get altered by `flipping' procedure. \\
 We now obtain the log divergent piece or the universal part of the partition function by expanding the integrand about $t=0$ and collect the $1/t$ term which is given by
\begin{align}
    \log\mathcal{Z}^{(4)}_v|_{\rm{univ}}=\frac{-1755 q^6-1680 q^4-35 q^2+6}{10080 q^5}-\frac{14}{45}.
\end{align}
Here $-\frac{14}{45}$ comes from the edge mode part of the partition function which is independent of the branching parameter $q$. 
 \subsection*{Conformal four-derivative vector field on $AdS_5\times S^1_q$}Hyperbolic cylinder $AdS_5\times S^1_q$ and branched sphere $S^6_q$ are conformally related. We express the partition function of the four-derivative vector gauge field on $AdS_5\times S^1_q$ in terms of integral over characters. This character integral method will help us to understand the relation between partition functions on these conformally related spaces.\\
 To evaluate the partition function one needs to fix the gauge properly. We follow the gauge fixing method introduced in \cite{Beccaria:2017dmw}. We decompose the field variables along the $S^1_q$ as well as on $AdS_5$ direction in the following way
 \begin{eqnarray}\label{gaugecondition}
A_{\mu}=\{A_{\tau},A^{\perp}_{i}+\nabla_{i}\phi\},\qquad\rm{where}\qquad \nabla_{i}A^{\perp i}=0
\end{eqnarray}
Here $i$, denotes the directions along $AdS_5$. Under the gauge transformation, the decomposed field variables transform in the following way
\begin{eqnarray}\label{gaugetrans1}
A_{\tau}\rightarrow A_{\tau}+\partial_{\tau}\epsilon,\qquad A^{\perp i}\rightarrow A^{\perp i},\qquad \phi\rightarrow \phi+\epsilon
\end{eqnarray}
Substituting the decomposed fields in the action we get
\begin{align}
\begin{split}
    S&=\int d^6x\sqrt{g}\left(\mathcal{L}_1+\mathcal{L}_2\right), \nonumber\\
    \mathcal{L}_1&=\left(\Delta_0^{\rm{AdS_5}} A^{\tau}-\partial_{\tau}\Delta^{\rm{AdS_5}}_0\phi\right)^2+\left(\Delta^{\rm{AdS_5}}_1 A^{\perp i}-\frac{R}{5}A^{\perp i}+\partial_{\tau}^2(A^{\perp i}+\nabla^{i}\phi)-\nabla^i A^{\tau}\right)^2,\nonumber\\
    \mathcal{L}_2&=\frac{R}{5}\left(\partial_{\tau}(A^{\perp i}+\nabla^{i}\phi)\right)^2.
\end{split}
\end{align}
Note that the action has the longitudinal mode along the $AdS_5$ direction.
$\Delta^{\rm{AdS_5}}_0$  and $\Delta^{\rm{AdS_5}}_1$ are the transverse spin-0 and spin-$1$ Laplacians on $AdS_5$ respectively. To obtain the action we have interchanged the covariant derivative and used the transverse gauge condition given in  \eqref{gaugecondition}. We know that
the change in field variables comes with the change in measure.
\begin{equation}\label{chgmea}
{\cal D} A_\mu =   [ {\rm d
et} ( - \Delta^{\rm{AdS_5}}_{0} )  ]^{1/2}  {\cal D} A_\tau  {\cal D} A_i^{\perp} {\cal D} \phi, 
\end{equation}
We treat the zero modes on the $S^1$ separately. Therefore
we can write the change of measure into that involving the zero modes on $S^1$ and the rest. 
\begin{eqnarray}
\label{chgmea2}
{\cal D} A_\mu &=&   \left[  ( {\rm d
et} ( - \Delta^{\rm{AdS_5}}_{0} )  )^{1/2}  {\cal D} A_\tau  {\cal D} A_i^{\perp} {\cal D} \phi
\right]_{n\neq 0}  \\ \nonumber 
& & \times  \left[ ( {\rm d
et} ( - \Delta^{\rm{AdS_5}}_{0} )  )^{1/2}  {\cal D} A_\tau  {\cal D} A_i^{\perp} {\cal D} \phi
\right ]_{n=0}.
\end{eqnarray}
\subsubsection*{Non-zero Kaluza-Klein modes}

We consider first the non-zero modes. 
From the gauge transformation given in (\ref{gaugetrans1}), one can fix the 
gauge $A_\tau =0$. Therefore we get a Fadeev-Popov determinant
\begin{align}
    \Delta_{\rm{FP}}=\det\partial_{\tau}.
\end{align}
 One introduces the Fadeev-Popov determinant along with the 
delta function which chooses this gauge slice 
\begin{equation}\label{gaugeslice}
{\cal G} = \delta (A_\tau)  {\rm det} ( \partial_\tau) .
\end{equation}
 Now we imagine setting $A_{\tau}=0$ in the action because of  the delta function in the partition function and separate out the action correspond to the transverse gauge field $A^{\perp i}$ and the longitudinal mode $\phi$ in the partition function
 \begin{align}
     \begin{split}
         \mathcal{L}_{A^{\perp}}^{n\neq 0}&=\left((-\partial_{\tau}^2-\Delta^{\rm{AdS_5}}_1-4)A^{\perp i}\right)^2-4\partial_{\tau}^2A^{\perp i}A^{\perp}_i\nonumber\\
       \mathcal{L}_{\phi}^{n\neq 0}&=\phi\partial_{\tau}^2(-\Delta^{\rm{AdS_5}}_0)(-\partial_{\tau}^2-\Delta^{\rm{AdS_5}}_0-4)\phi   
     \end{split}.
 \end{align}
 Therefore we can write the partition function for the non-zero Kaluza-Klein modes
 \begin{align}
{\cal Z}^{(1)}_{\rm{hyp},n\neq 0} =& \int \left[  {\cal D} A_i^\perp \;  (  {\rm d
et} ( - \Delta^{\rm{AdS_5}}_{0} )^{1/2}  {\rm det} ( \partial_\tau)   \right]_{n\neq 0}
\times \nonumber\\ 
&\left[  \frac{1}{  {\rm det} ( \partial_\tau) ( {\rm d
et} ( - \Delta^{\rm{AdS_5}}_0 ) )^{1/2}{\rm d
et} (- \partial_{\tau}^2-\Delta^{\rm{AdS_5}}_0-4) ) }  \right]_{n\neq 0}  
  \times 
\exp \left\{ - \frac{1}{2} \int d^6 x  \sqrt{g}    \mathcal{L}_{A^{\perp}}^{n\neq 0}\right\}.
\end{align}
Here we have integrated the $A_{\tau}$ first using the gauge slice given in \eqref{gaugeslice}.
Note that determinants coming from the change in measure and the Fadeev-Popov determinant cancel one part of the determinant from the longitudinal mode. Therefore integrating over the transverse gauge field, the partition function  becomes
\begin{align}\label{nonzeromodepartition}
   {\cal Z}^{(1)}_{\rm{hyp},n\neq 0}&=\left(\frac{1}{\det(\left(-\partial_{\tau}^2-\Delta^{\rm{AdS_5}}_1-4)+4\partial_{\tau}^2\right)_{n\neq 0}\det\left(- \partial_{\tau}^2-\Delta^{\rm{AdS_5}}_0-4\right)_{n\neq 0}}\right)^{\frac{1}{2}}.
\end{align}
\subsection*{Kaluza-Klein zero modes:}The action for the zero modes becomes simple
\begin{align}\label{veczeromod}
    S_{n=0}&=\int\sqrt{g}d^6x \left(A_{\tau\, n=0}(-\Delta^{\rm{AdS_5}}_0)(-\Delta^{\rm{AdS_5}}_0-4)A_{\tau\, n=0}+\left((\Delta^{\rm{AdS_5}}_1-4) A_{i\, n=0} ^\perp\right)^2\right).
\end{align}
Here the Kaluza-Klein zero mode of $A_\tau$ 
 along the $S^1$ is gauge invariant. From (\ref{gaugetrans1}) it can be seen 
that it transforms as 
\begin{equation}
A_{\tau, \, n=0} \rightarrow A_{\tau, \, n=0}.
\end{equation}
The action (\ref{veczeromod})  is independent of  the longitudinal mode
$\phi$ and therefore is gauge invariant under a constant shift
\begin{equation}
\phi_{n=0}  \rightarrow \phi_{n=0} + \epsilon_{n=0}.
\end{equation}
We fix this gauge by restricting to the slice  
\begin{equation}
{\cal G} = \delta ( \phi_{n=0} ) .
\end{equation}
The Fadeev-Popov
determinant for this choice of gauge slice is unity.
\begin{eqnarray}\label{zeromodepartition}
{\cal Z}^{(1)}_{\rm{hyp},n= 0} &=& \int  \left[ {\cal D}A^\perp_{i} {\cal  D}A_\tau  {\rm det} ( -\Delta^{\rm{AdS_5}}_{(0)} )^{\frac{1}{2}} \right]_{n=0} 
\\ \nonumber
& & \times
\exp\left\{ - \frac{1}{2} \int d^6 x  \sqrt{g}  \left(
 A_{i\, n=0} ^\perp (  -\Delta^{\rm{AdS_5}}_1 - 4 ) 
 A^{\perp\,  i } _{n=0} 
 + A_{\tau\, n =0} ( - \Delta^{\rm{AdS_5}}_0 )(-\Delta^{\rm{AdS_5}}_0-4) A_{\tau\, n=0}   \right ) \right\} \\ \nonumber
 &=& \left[ \frac{1}{ {\rm det} ( -\Delta^{\rm{AdS_5}}_1 -4 ) {\rm det}(-\Delta^{\rm{AdS_5}}_0-4) } \right]^{\frac{1}{2}}.
\end{eqnarray}
It is important to note that for the $n=0$ modes, the Jacobian from the change in measure is cancelled by the integration of 
$A_\tau$ which is gauge invariant. Therefore combining \eqref{nonzeromodepartition} and \eqref{zeromodepartition} we obtain
\begin{align}\label{fullpartition}
   {\cal Z}^{(1)}_{\rm{hyp},n= 0}&=\left(\frac{1}{\det(\left(-\partial_{\tau}^2-\Delta^{\rm{AdS_5}}_1-4)+4\partial_{\tau}^2\right)\det\left(- \partial_{\tau}^2-\Delta^{\rm{AdS_5}}_0-4\right)}\right)^{\frac{1}{2}}.
\end{align}
We now factorise the the operator $\mathcal{W}^{(4)}_{AdS_5\times S^1}=\det\left((-\partial_{\tau}^2-\Delta^{\rm{AdS_5}}_1-4)+4\partial_{\tau}^2\right)$.
\begin{align}
    \mathcal{W}^{(4)}_{AdS_5\times S^1}&=\det(\left(-\partial_{\tau}^2-\Delta^{\rm{AdS_5}}_1-4)+4\partial_{\tau}^2\right)\nonumber\\
    &=\left(-\partial_{\tau}^2+(\sqrt{(-\Delta^{\rm{AdS_5}}_1-4)}+i)^2\right)\left(-\partial_{\tau}^2+(\sqrt{(-\Delta^{\rm{AdS_5}}_1-4)}-i)^2\right).
\end{align}
 $\Delta_1$ is the transverse spin-1 Laplacian on $AdS_5$ satisfying the eigen value equation
  \begin{equation}
\Delta^{\rm{AdS_5}}_1\psi^{\{\lambda, u\}} = -\left[  \lambda^2 +   4 \right] 
 \psi_\lambda^{\{\lambda, u\}} ,
 \end{equation}
  $\psi^{ \{\lambda, u\} }$  are the corresponding eigen functions, $\{u\}$ labels other quantum numbers
 on $AdS_5$. 
Using these eigen values  and the Kaluza-Klein decomposition of the partition function 
we obtain
\begin{align}
\frac{1}{2}\log\mathcal{W}^{(4)}_{AdS_5\times S^1_q}
   =
 \frac{1}{4}  \sum_{n=-\infty}^\infty \int_{-\infty}^\infty d\lambda \mu_{1}^{(5)}(\lambda) 
   \left( \log \left( \frac{n^2}{q^2} +(\lambda+i)^2 \right)+\log \left( \frac{n^2}{q^2} +(\lambda-i)^2
 \right)\right).
\end{align}
Now we follow the steps from \eqref{feq1} to \eqref{lastustep} and obtain
\begin{align}
  \frac{1}{2}  \log \mathcal{W}^{(4)}_{AdS_5\times S^1_q}&=\frac{\log R}{4\pi i}\int_{\rm{C_{Odd}}}du\int_0^{\infty}\frac{d\tau}{\sqrt{4\pi\tau^3}} \sum_{n=0}^{\infty}e^{-\frac{\epsilon^2+u^2}{4\tau}}e^{-\tau\frac{(n)^2}{q^2}}\nonumber\\
   & \times\int_{-\infty}^{\infty}d\lambda  \mu^{(5)}_1 (\lambda)(e^{i(\lambda+i)u}+e^{i(\lambda-i)u})\nonumber\\
   &=\frac{\log R}{4\pi i}\int_{\rm{C_{Odd}}}du\int_0^{\infty}\frac{d\tau}{\sqrt{4\pi\tau^3}} \sum_{n=0}^{\infty}e^{-\frac{\epsilon^2+u^2}{4\tau}}e^{-\tau\frac{(n)^2}{q^2}}W_1^{(5)}( u)(e^u+e^{-u}).
\end{align}
$W_1^{(5)}( u)$ is the Fourier transform of the  Plancherel measure of transverse spin-1 Laplacian on $AdS_5$ which is given by \cite{Sun:2020ame}
\begin{align}
   W_1^{(5)}( u)= \frac{(\exp (-u)+1) (\exp (-3 u)-6 \exp (-2 u)+\exp (-u))}{(1-\exp (-u))^5}.
\end{align}
Now we  perform the  integral over $\tau$ and sum over the Kaluza-Klein modes to obtain
\begin{align}\label{firstpart}
   \frac{1}{2} \log \mathcal{W}^{(4)}_{AdS_5\times S^1_q}&=\frac{\log R}{2\pi i}\int_{\rm{C_{Odd}}}\frac{du}{2u}\frac{1+e^{-\frac{u}{q}}}{1-e^{-\frac{u}{q}}}\frac{(1+e^{-u})(e^{-3u}-6e^{-2u}+e^{-u})}{(1-e^{-u})^5}(e^u+e^{-u}).
\end{align}
We have another determinant of conformal scalar on $AdS_5\times S^1$ in \eqref{fullpartition}. Note that the mass $m_1^2=-4$ saturates the Breitenlohner-Freedman bound of spin-0 operator on $AdS_5$. Therefore we can write
\begin{align}
  \frac{1}{2}  \log{\det\left(- \partial_{\tau}^2-\Delta_0-4\right)}&= \frac{1}{4}  \sum_{n=-\infty}^\infty \int_{-\infty}^\infty d\lambda \mu_{0}^{(5)}(\lambda) 
   \log \left( \lambda^2+\frac{n^2}{q^2}\right).
\end{align}
We follow the same procedure to obtain
\begin{align}\label{secondpart}
   \frac{1}{2}  \log{\det\left(- \partial_{\tau}^2-\Delta_0-4\right)}&=\frac{\log R}{2\pi i}\int_{\rm{C_{Odd}}}\frac{du}{2u}W_0^{(5)}\frac{1+e^{-\frac{u}{q}}}{1-e^{-\frac{u}{q}}}\nonumber\\
     &=\frac{\log R}{2\pi i}\int_{\rm{C_{Odd}}}\frac{du}{2u}\frac{1+e^{-\frac{u}{q}}}{1-e^{-\frac{u}{q}}}\left(-\frac{e^{-2 u} \left(e^{-u}+1\right)}{\left(1-e^{-u}\right)^5}\right).
\end{align}
We combine the \eqref{firstpart} and the \eqref{secondpart} and obtain
\begin{align}
    \log\mathcal{Z}^{(1)}_{\rm{hyp}}&=\frac{\log R}{2\pi i}\int_{\rm{C_{Odd}}}\frac{du}{2u}\frac{1+e^{-\frac{u}{q}}}{1-e^{-\frac{u}{q}}}\left(5\frac{(e^{-u}+e^{-4u})}{(1-e^{-u})^5}+5\frac{(e^{-2u}+e^{-3u})}{(1-e^{-u})^5}-\frac{(1+e^{-5u})}{(1-e^{-u})^5}\right).
\end{align} 
Note that the partition function is precisely the integral over `naive' bulk-character of four-derivative conformal vector on $S^6_q$ given in \eqref{spherevector}. Therefore the partition function of the conformal four-derivative vector field on hyperbolic cylinder misses out the edge character. It follows the pattern which was noticed for the Maxwell field in four-dimensions. The `extractable' part of the entanglement entropy of free Maxwell field precisely agrees with the hyperbolic cylinder computation. In \cite{Beccaria:2017dmw} the entanglement entropy of the higher derivative vector gauge theory is computed from the free energy of the hyperbolic cylinder. Following that we obtain free energy and the universal part of the R\'{e}nyi entropy
\begin{align}
    \mathcal{F}^{(1)}_{\rm{hyp},q}=-\frac{-1755 q^6-1680 q^4-35 q^2+6}{10080 q^5},\qquad S^{(1)}_{\rm{hyp},q}=-\frac{(q+1) \left(1709 q^4+29 q^2-6\right)}{10080 q^5}.
\end{align}
The R\'{e}nyi entropy precisely agrees with \cite{Beccaria:2017dmw} but free energy differs by a  term linear in $q$. The reason is that we have included the zero-mode to cast the partition function in terms of `naive' characters. Note that the powers which are linear in $q$ do not affect the R\'{e}nyi entropy which is therefore UV-finite.

The partition function of the higher derivative conformal vector field on the hyperbolic cylinder does not capture the edge mode and one can not obtain the anomaly coefficients directly from it.
Therefore the $a$-anomaly coefficient can be obtained from the partition function on $S^6_q$
\begin{align}
 \frac{- \log\mathcal{Z}^{(4)}_v|_{q=1}-\partial_q\log\mathcal{Z}^{(4)}_v|_{q=1}}{96}=\frac{275}{8\times 7!}.
\end{align}
This $a$-anomaly coefficient precisely agrees with \cite{Beccaria:2015uta}. Note that in \cite{Beccaria:2017dmw} it was mentioned that to get the correct $a$-anomaly coefficient one needs to shift the free energy by $-\frac{14}{45}$. Here we clearly observe that the shift is coming from the edge mode part of the partition function. Therefore the partition function of the four-derivative conformal vector on $S^6_q$ correctly reproduces the $a$-anomaly coefficients but the partition function on the hyperbolic cylinder gives the entanglement entropy. It will be interesting if one can directly compute the entanglement entropy by using radial quantization procedure \cite{Benedetti:2019uej} or show that the `extractable' part of the entanglement entropy \cite{Soni:2016ogt} comes from the hyperbolic cylinder. We keep these computations for future purposes.\\
We evaluate the conformal dimension of the twist operator from the free energy of the four-derivative vector field on hyperbolic cylinder.
\begin{align}\label{hqhvec}
    h_q&=\frac{q}{5\rm{Vol}(AdS_5)}(\partial_q\mathcal{F}^{(1)}_{\rm{hyp},q}|_{q=1}-\partial_q\mathcal{F}_{\rm{hyp},q})\nonumber\\
    &=\frac{-117 q^6+112 q^4+7 q^2-2}{3360 \pi ^2 q^5}.
\end{align}
The regularised volume of $AdS_5$ can be obtained from \eqref{volads}.
The first and the second derivative of $h_q$ with respect to $q$ at $q=1$ can be expressed in terms of the parameters relating two and three point functions of the stress tensor. 
\begin{align}\label{hqdervec}
    h_q^{\prime}|_{q=1}=-\frac{1}{14 \pi ^2},\qquad h_q^{\prime\prime}|_{q=1}=\frac{31}{420 \pi ^2}.
\end{align}These numbers will be useful to study the causality constraints of the four-derivative conformal vector field  will be  discussed in section \eqref{causality}.
\section{Hofman-Maldacena variables and causality bound}\label{causality}
Higher derivative kinetic term in the action implies negative residue in the propagator which indicates a negative norm state \cite{Donoghue:2017fvm}. Therefore one expects the theories with the higher derivative kinetic terms are non-unitary. In this section, we use the positivity of the energy flux or average null energy condition to demonstrate that  these theories are indeed non-unitary. 

The co-dimension two twist operator essentially captures the energy density across the entangling surface \cite{Hung:2014npa}. The energy density can be evaluated in a thermal ensemble on the hyperbolic cylinder. Therefore the first and second derivative of the conformal dimension of the twist operator with respect to $q$ at $q=1$ can be expressed in terms the two and three point functions of the stress tensor because one can think $q$ as inverse temperature  in a thermal ensemble. In conformal field theory  the two and three point functions of the stress tensor are characterized by the parameters $a$, $b$ and $c$ \cite{Osborn:1993cr}
 and these parameters are expected to obey the following relations
\begin{eqnarray}\label{hqabc}
& & h_q^\prime|_{q=1}=\frac{\pi ^{\frac{d+3}{2}} C_T}{2^{d-3} d \left(d^2-1\right) \Gamma \left(\frac{d-1}{2}\right)} , \quad
C_T =\frac{\left(8  \pi ^{d/2}\right) (a (d-2) (d+3)-2 b-c (d+1))}{d (d+2) \Gamma \left(\frac{d}{2}\right)},
\nonumber \\
& &  \nonumber
h_q^{\prime\prime} |_{q=1} =-\frac{16 \pi ^{d+1} }{d^2 \Gamma (d+3)}
 \left[2 a \left(3 d^2-3 d-4\right) (d-2)-2 b d (d-1)-c (3 d-4) (d+1)\right],
 \\
\end{eqnarray}
where $C_T$ is the central charge of the theory and $a, b, c$ are the parameters determining the $3$ point functions of the stress tensor \cite{Osborn:1993cr}. Since we have two relations we can not determine all the parameters from the relations given in \eqref{hqabc}. Here we wish to mention that  $h_q^{\prime\prime}|_{q=1}$ does not match when one substitutes the values of the parameters $a$, $b$ and $c$ for the conformally coupled real scalar and the expression was modified for the conformal scalars \cite{Herzog:2016bhv}. But $h_q^{\prime} |_{q=1}$ and $h_q^{\prime\prime} |_{q=1}$ agree with explicit verification for conformal fermions and conformal $p$-form fields \cite{David:2020mls}. In this case, we assume that the relations hold true for the higher derivative conformal fields. 

 Hofman-Maldacena variables $t_2$ and $t_4$  were constructed from ratio of the combinations of the three parameters $a$, $b$ and $c$ to characterize the positive energy of the flux \cite{Hofman:2008ar}. The relations are known for arbitrary dimensions \cite{Buchel:2009sk}
\begin{align}
    \begin{split}
     t_2&=\frac{2 (d+1) (a (d-1) ((d+8) d+4)-d (-3 b d+2 c d+c))}{d (a (d-2) (d+3)-2 b-c (d+1))},\\
     t_4&=\frac{(d+1) (d+2) \left(3 a \left(-2 d^2+d+1\right)+d (-2 b d+c d+c)\right)}{d (a (d-2) (d+3)-2 b-c (d+1))}.
    \end{split}
\end{align}
The positivity of the energy flux constraints can be related to the causality and the unitarity of the conformal field theory \cite{Kulaxizi:2010jt,Hartman:2015lfa,Hofman:2016awc} which puts the bound on the parameters $t_2$ and $t_4$ for conformal field theories in $d>3$ dimensions and they are given by \cite{Camanho:2009vw} 
\begin{align}\label{bounds}
    \begin{split}
    g_1&=1 -\frac{t_2}{d-1}-\frac{2t_4}{(d+1) (d-1)}\geq 0,\\
  g_2 &=1 -\frac{t_2}{d-1}-\frac{2 t_4}{(d-1) (d+1)}+\frac{t_2}{2}\geq 0,\\
 g_3&=1+  \frac{(d-2) (t_2+t_4)}{d-1}-\frac{t_2}{d-1}-\frac{2 t_4}{(d-1) (d+1)}\geq 0.
    \end{split}
\end{align}
Therefore these bounds on the parameters imply that any  field theory which obey the causality constraint lie in the region bounded by the three lines given in \eqref{bounds}. For $d=6$ this is shown by a shaded triangle and any field theory in $d=6$ satisfying the causality constraint will lie in the shaded region.
\begin{figure}[htb]
\begin{center}
   \includegraphics[scale=0.4]{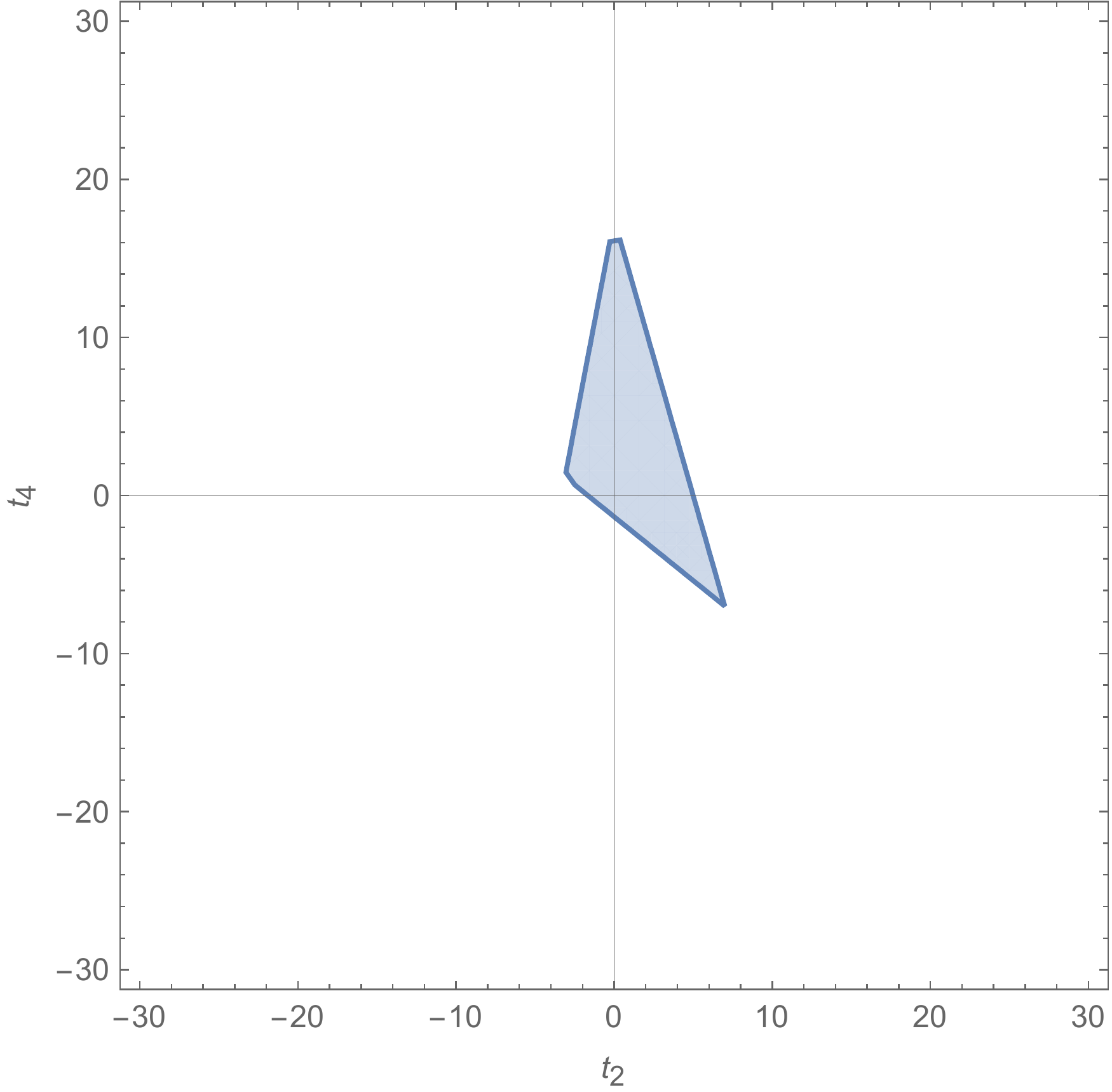}
   \caption{The allowed region in $t_2$ and $t_4$ in $d=6$ for Weyl-invariant field theories satisfying causality constraints}
\end{center}
\end{figure} 
From \eqref{hqabc} we find two relations of the three parameters $a$, $b$ and $c$. Therefore we can not fully determine the values of the parameters. However one can find a linear relation of the parameters $t_2$ and $t_4$ which will be a straight line in $t_2-t_4$ plane and the theory will be a point on the line. If the line never touches the triangle one can conclude that the theory does not satisfy the causality constrain. We verify it first for the four-derivative conformal scalar in $d=4$ dimensions.
\begin{figure}[htb]
\centering
\begin{subfigure}{.5\textwidth}
  \centering
  \includegraphics[width=1\linewidth]{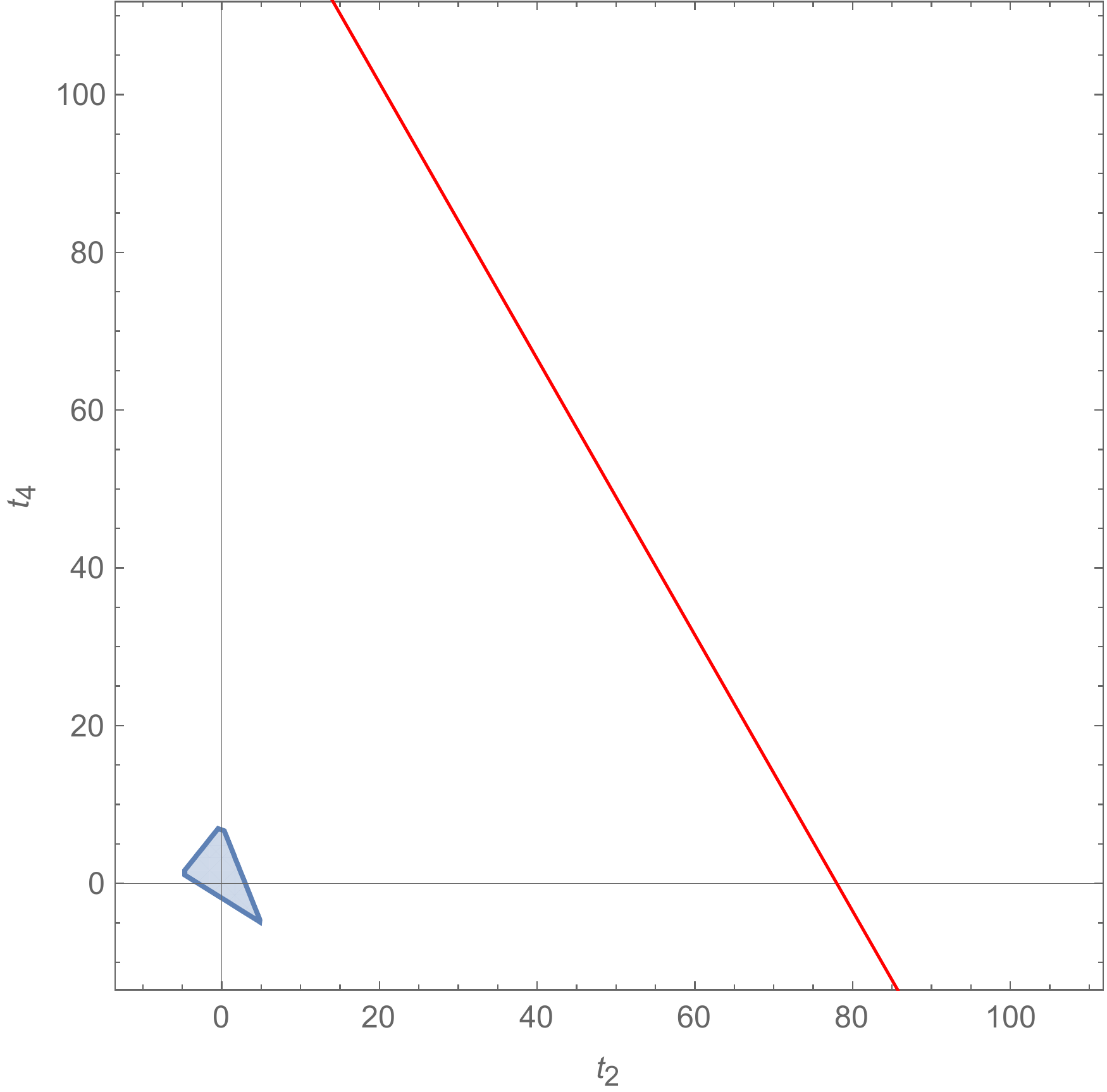}
  \caption{$4$-derivative conformal scalar in $d=4$.}
  \label{fig:sub12}
\end{subfigure}%
\begin{subfigure}{.5\textwidth}
  \centering
  \includegraphics[width=1\linewidth]{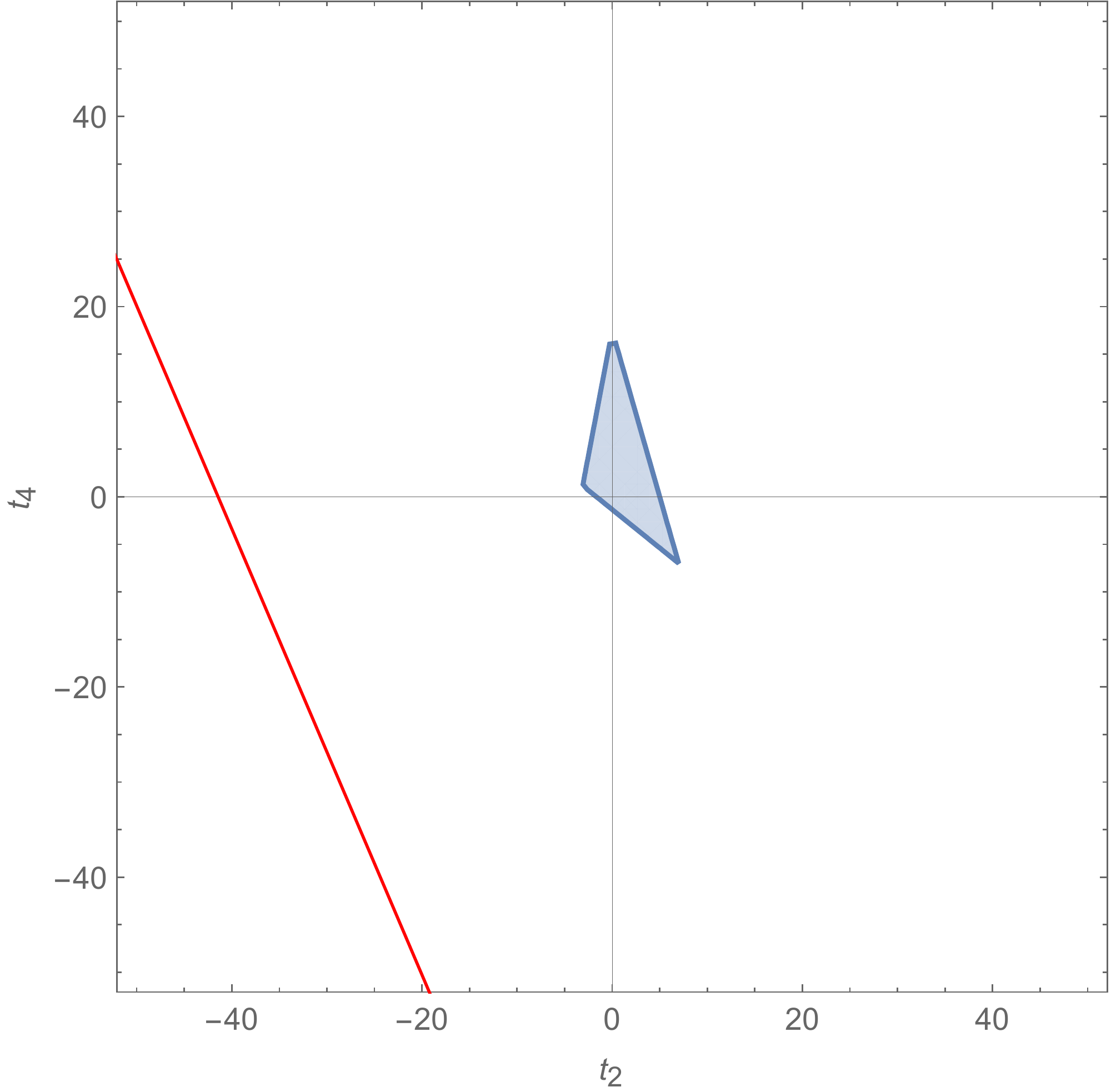}
  \caption{$6$-derivative conformal scalar in $d=6$.}
  \label{fig:sub22}
\end{subfigure}
\label{fig:test1}
\caption{Higher derivative conformal scalar field is constrained to lie on the red line.}
\end{figure}
Here blue-shaded region denotes all the  theories satisfying causality constraint  in $d=4$ dimensions. Note that $4$-derivative Weyl-invariant scalar is a point in the $t_2-t_4$ plane but it is now constrained to lie on a straight line given by
\begin{align}
    t_2+\frac{2}{7}  (2 t_4-273)=0.
\end{align}
The intercepts of the straight line are given by $t_2=78.0$ and $t_4=136.5$ and it never touches the bounded region. Therefore any point on the straight line will not lie inside the bounded region. We have checked this in arbitrary dimensions and therefore conclude that the $4$-derivative Weyl-invariant scalar field does not satisfy the causality constraint.\\
Similarly for $6$-derivative conformal scalar we obtain linear relation of Hofman-Maldacena variables from \eqref{6hder} and the relation is given by
\begin{align}
    t_2-\frac{1}{462} (-197 t_4-19145)=0.
\end{align}
This represents a straight line in $t_2-t_4$-plane and the intercepts are given by $t_2=-41.4394$, $t_4=-97.1827$. We observe that the straight line never touches the region of causality which is a triangular region. Therefore the $6$-derivative conformal scalar in $d=6$ dimension which is a point on the straight line can never lie inside the region of causality. So the $6$-derivative conformal scalar in $d=6$ dimension also does not obey the causality constraint.
We repeat the same analysis for the $3$-derivative Weyl invariant spin-half field. The weight of the twist operator of the $3$-derivative conformal spin-half field is given in \eqref{hqfer} and the first and second derivative of it with respect to $q$ at $q=1$ are given by
\begin{align}
    h_q=-\frac{29 q^4-50 q^2+21}{2880 \pi  q^3},\qquad h_q^{\prime}=-\frac{1}{180 \pi },\qquad h_q^{\prime\prime}=-\frac{19}{360 \pi }.
\end{align}
Given the values of $h_q^{\prime}|_{q=1}$ and $h_q^{\prime\prime}|_{q=1}$ we find a linear relation of the variable $t_2$ and $t_4$ which is a straight line on the plane. The equation of the straight line is given by
\begin{align}
    \ell _f=t_4+\frac{1}{4} \left(7 t_2+12414\right)=0.
\end{align}
The intercepts are given by $t_2=-1773.43$ and $t_4=-3103.5$. Therefore it never crosses the region of causality which is the the bounded region by $g_1$, $g_2$ and $g_3$ shown in \eqref{fig:fer1}. So Weyl covariant $3$-derivative fermionic field also does not obey the causality bound.

Now we examine the 4-derivative Weyl-covariant vector field in $d=6$ dimensions.
We obtain the conformal dimension of the twist operator and the first and second derivative of it with respect to $q$ at $q=1$ which are given in \eqref{hqhvec} and \eqref{hqdervec} respectively.
We substitute these values in equations \eqref{hqabc} and use \eqref{bounds} to obtain a relation between Hofman-Maldacena variables
\begin{align}
 \ell_g= t_4+  \frac{231}{197} \left(2 t_2-1105\right)=0.
\end{align}
Therefore it is again a straight line in $t_2-t_4$ plane.
\begin{figure}
\centering
\begin{subfigure}{.5\textwidth}
  \centering
  \includegraphics[width=1\linewidth]{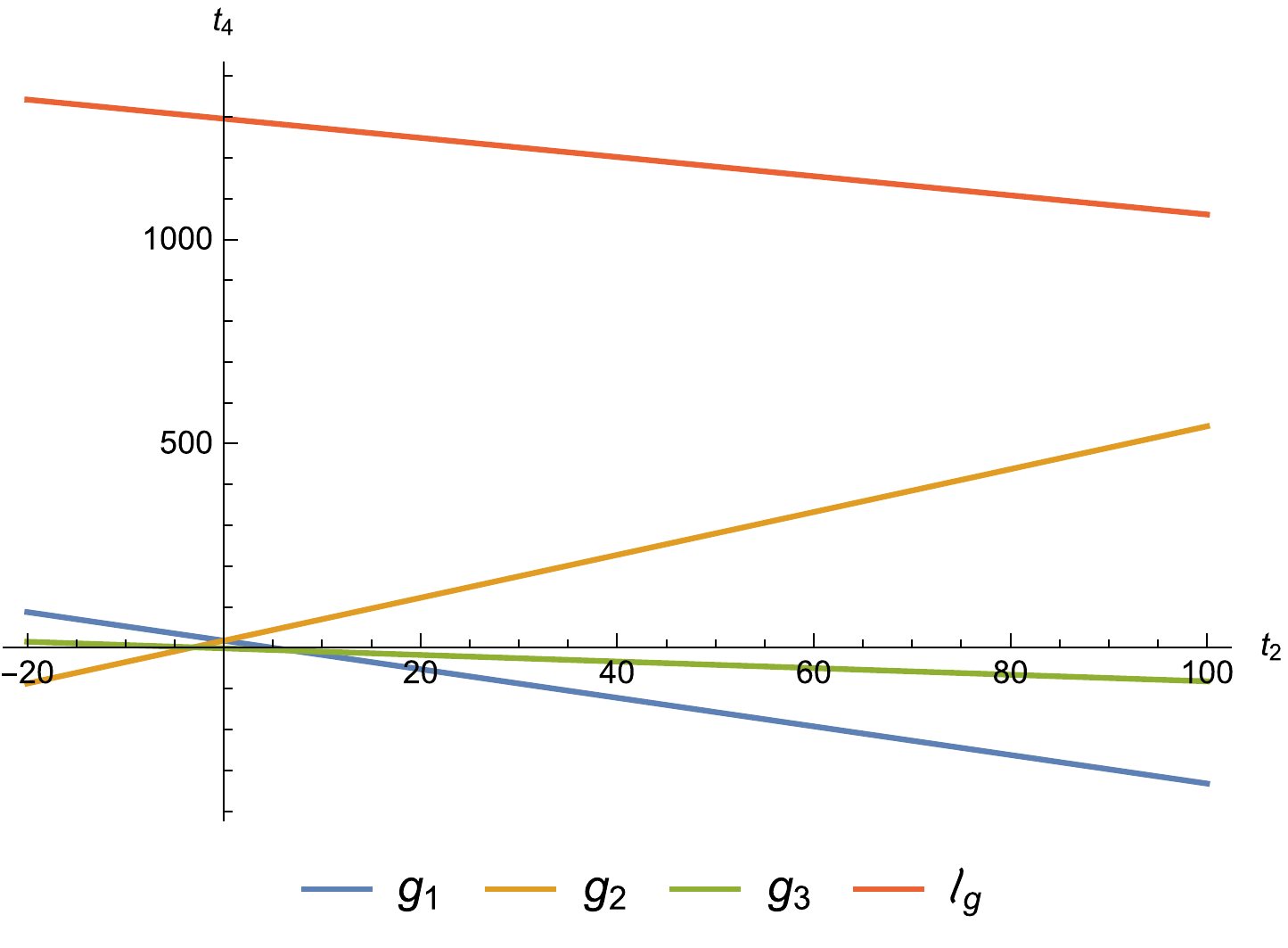}
  \caption{$4$-derivative conformal vector in $d=6$.}
  \label{fig:vec1}
\end{subfigure}%
\begin{subfigure}{.5\textwidth}
  \centering
  \includegraphics[width=1\linewidth]{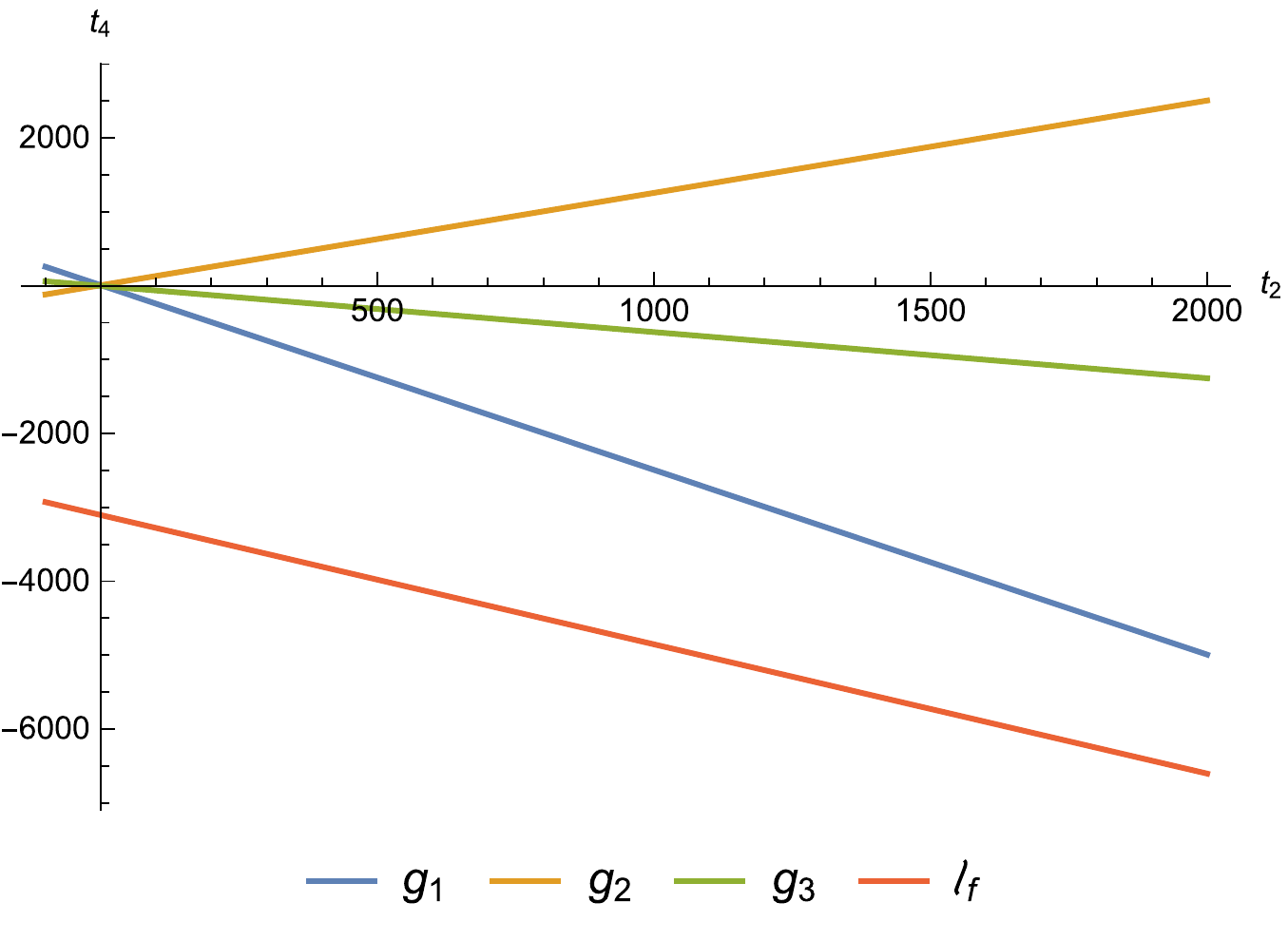}
  \caption{$3$-derivative conformal fermion in $d=4$.}
  \label{fig:fer1}
\end{subfigure}
\label{fig:test}
\caption{The region of causality is the small triangle where $g_1$, $g_2$ and $g_3$ meet near the origin.}
\end{figure}
In the figure \eqref{fig:vec1}, $g_1$, $g_2$ and $g_3$ are the lines given in \eqref{bounds}. $\ell_g$  represents the straight line on which the $4$-derivative Weyl-covariant vector gauge theory lies. Note that the region of causality bound will be the small triangular region created by the three lines $g_1$, $g_2$ and $g_3$ and the red line does not pass through the triangular region. The intercepts of the red line $\ell_g$ on the axes are $t_2=552.5$ and $t_4=1295.71$. Therefore it never cuts the triangular region at any point. So we conclude that $4$-derivative Weyl-covariant vector field does not obey the causality constraints.\\
We observe that the higher derivative conformal theories do not lie inside the region of causality because these theories do not satisfy the average null energy condition. Therefore we conclude that the higher derivative conformal fields are non-unitary.

\section{Conclusions}
In this paper, we have applied the formalism developed in \cite{Anninos:2020hfj, Sun:2020ame, David:2021wrw} to express the one loop partition functions in terms of integrals over characters of higher derivative conformal fields. The integral representation is useful to compare the partition functions on conformally related spaces. The partition function of higher derivative conformal scalar and fermionic fields are identical on conformally related spaces but the partition function of $4$-derivative vector gauge field misses out the edge mode on the hyperbolic cylinder. This follows the pattern which was observed for conformal $p$-forms on hyperbolic cylinder\cite{David:2021wrw}. It was also observed that the partition function of the free Maxwell field on the hyperbolic cylinder gives the `extractable' part of the entanglement entropy. Therefore it will be interesting to compute the `extractable' part of entanglement entropy for $4$-derivative vector gauge field in $d=6$ dimensions and verify with the hyperbolic cylinder computation.

One can also investigate the conformal higher spin fields \cite{Beccaria:2017lcz} on conformally related spaces. This method of obtaining one-loop partition functions will be helpful to check whether the hyperbolic cylinder misses out on the `edge' mode or not. One can again repeat the computations developed in \cite{Soni:2016ogt} to understand the `extractable' entanglement entropy of the conformal higher spin fields. It will be interesting to perform an alternative analysis to evaluate the entanglement entropy which is based on the radial quantization of the fields on sphere \cite{Benedetti:2019uej} and check with the computation of free energy on the sphere or hyperbolic cylinder.

We have shown that higher derivative conformal theories do not obey the causality bound and therefore it supports the earlier observations regarding the non-unitarity of these theories. Here we could not determine all the parameters $a$, $b$, and $c$  but we found a linear relationship which is a straight line in $t_2-t_4$ plane and the theory is a point on the straight line. We observe that the line never touches the region of causality bound. It will be interesting to compute two and three-point functions of the stress tensors to determine all the parameters exactly to show that the theory lies outside of the region of causality bound.
 \acknowledgments
 The author wishes to thank Justin  David for the fruitful discussions, suggestions on the manuscript, and encouragements. He also thanks Sarthak Duary for discussions.
\bibliographystyle{JHEP}
\bibliography{References.bib}
\end{document}